\documentclass{aastex}
\usepackage{emulateapj5, apjfonts, psfig, epsfig}
\begin{document}
\slugcomment{AJ, in press}

\title{M75, a Globular Cluster with a Trimodal Horizontal Branch.\\ 
       I.~Color-Magnitude Diagram}

\shorttitle{The Trimodal HB of M75. I. CMD analysis}
\shortauthors{Catelan  et al.}

\received{}
\revised{}
\accepted{}

\author{M.~Catelan}
\affil{Pontificia Universidad Cat\'olica de Chile, 
       Departamento de Astronom\'\i a y Astrof\'\i sica, 
       Av. Vicu\~{n}a Mackenna 4860, 6904411 Macul, 
       Santiago, Chile 
}
\email{mcatelan@astro.puc.cl} 

\author{J.~Borissova}
\affil{Institute of Astronomy, Bulgarian Academy of Sciences and Isaac 
       Newton Institute of Chile Bulgarian Branch,
       72~Tsarigradsko chauss\`ee, BG\,--\,1784 Sofia, Bulgaria}
\email{jura@haemimont.bg}

\author{F.~R.~Ferraro}
\affil{Osservatorio Astronomico di Bologna,
       Via Ranzani 1, I-40127 Bologna, Italy}
\email{ferraro@apache.bo.astro.it} 

\author{T.~M.~Corwin}
\affil{Department of Physics, University of North Carolina at Charlotte, 
       Charlotte, NC 28223}
\email{mcorwin@uncc.edu} 

\author{H.~A.~Smith}
\affil{Department of Physics and Astronomy, Michigan State University, 
       East Lansing, MI 48824-1116}
\email{smith@pa.msu.edu} 

\and
\author{R.~Kurtev}
\affil{Department of Astronomy, Sofia University and Isaac Newton Institute 
       of Chile Bulgarian Branch, James Bourchier Ave. 5, BG\,--\,1164 Sofia,
       Bulgaria} 
\email{kurtev@phys.uni-sofia.bg}

\begin{abstract}
    Deep $UBVI$ photometry for a large field covering the distant  
    globular cluster M75 (NGC~6864) is presented. We confirm a previous 
    suggestion (Catelan et al. 1998a) that M75 possesses a bimodal 
    horizontal branch (HB) bearing striking resemblance to the well-known 
    case of NGC~1851. In addition, we detect a third, smaller grouping of  
    stars on the M75 blue tail, separated from the bulk of the blue HB stars  
    by a gap spanning about 0.5~mag in $V$. Such a group of stars may 
    correspond to the upper part of a very extended, though thinly 
    populated, blue tail. Thus M75 appears to have a  {\em trimodal} HB. 
    The presence of the  ``Grundahl jump" is verified using the broadband 
    $U$ filter. We explore the color-magnitude diagram of M75  
    with the purpose of deriving the cluster's fundamental parameters, and 
    find a metallicity of ${\rm [Fe/H]} = -1.03 \pm 0.17$~dex and 
    $-1.24 \pm 0.21$ in the Carretta \& Gratton (1997) and Zinn \& West 
    (1984) scales, respectively.  We discuss earlier suggestions that the cluster  
    has an anomalously low ratio of bright red giants to HB stars. 
    A differential age analysis with respect 
    to NGC~1851 suggests that the two clusters are essentially coeval.
\end{abstract} 


\section{Introduction}
Adequate interpretation of the color-magnitude diagrams (CMDs) of globular clusters 
(GCs) is important for many astrophysical and cosmological reasons. The horizontal 
branch (HB) evolutionary phase plays a particularly significant role in that 
regard. Because the observational properties of HB stars are extremely sensitive 
to the physical parameters of GCs, HB stars can impose invaluable constraints on 
the initial conditions that characterized the Galaxy in its infancy. 

Metallicity, [Fe/H], is the  ``first parameter" influencing HB morphology: Metal-rich 
clusters have, in general, more red HB stars than blue HB stars and {\em vice versa} 
for the metal-poor GCs. The existence of GCs which have the same [Fe/H] but different 
HB morphologies indicates that one or more second parameters exists. Despite the 
fact that the problem has received much attention since it was first discovered (Sandage 
\& Wildey 1967), the second parameter(s) responsible for the spread in 
HB types at a given [Fe/H] has (have) not been firmly established. Understanding 
the second parameter is arguably one of the most important tasks that must be 
accomplished before the age and formation history of the Galaxy can be reliably 
determined (e.g., Searle \& Zinn 1978; van den Bergh 1993; Zinn 1993; 
Lee, Demarque, \& Zinn 1994; Majewski 1994). 

Much recent debate has focused on whether the predominant second parameter is age 
(e.g., Lee et al. 1994; Stetson, VandenBerg, \& Bolte 1996; 
Sarajedini, Chaboyer, \& Demarque 1997; Catelan 2000; Catelan, Ferraro, \& Rood 
2001b; Catelan et al. 2001a; 
VandenBerg 2000), but mass loss, helium and $\alpha$-element abundances, 
rotation, deep mixing, binary interactions, core concentration, and even planetary 
systems have been suggested as second parameters. It is likely that the second 
parameter phenomenon is in fact a combination of several individual quantities 
and phenomena related to the formation and chemodynamical evolution of each 
individual star cluster, implying that the several individual second-parameter 
candidates have to be  ``weighted" on a case-by-case basis (e.g., Fusi Pecci et al. 
1993; Fusi Pecci \& Bellazzini 1997; 
Rich et al. 1997; Sweigart 1997a, 1997b; Kraft et al. 1998; 
Sweigart \& Catelan 1998; Soker 1998;  Ferraro et al. 1999a; Catelan 2000; 
Soker \& Harpaz 2000; Catelan et al. 2001a, 2001b). 
In this context, it has been extensively argued that an understanding of GCs 
presenting the second-parameter syndrome {\em internally}---i.e., those with 
{\em bimodal} HBs---is of paramount importance for understanding the nature of the 
second-parameter phenomenon as a whole (e.g., Rood et al. 1993; Stetson et al. 
1996; Borissova et al. 1997; Catelan et al. 1998a; Fusi Pecci \& Bellazzini 1997; 
Walker 1998). 

Unfortunately, the current list of GCs known to have a bimodal 
HB is not large; probably only NGC~1851, NGC~2808 and NGC~6229 are universally 
recognized as  ``bona fide" bimodal-HB clusters, following the definition of the 
term provided by Catelan et al. (1998a). However, as discussed by Catelan et al., 
there are several other GCs whose poorly investigated CMDs, once improved
on the basis of modern CCD photometry, {\em may} well move
them into the  ``bimodal HB" category. Outstanding among those is M75 (NGC~6864). 

M75 is a moderately metal-rich (${\rm [Fe/H]} \simeq -1.3$ in the Zinn \& West 
1984, hereafter ZW84, 
scale; Harris 1996), dense \{$\log[\rho_0/(M_{\odot}{\rm pc}^{-3})] = 4.9$; 
Pryor \& Meylan 1993\} and distant ($R_{\odot} \simeq 19$~kpc) halo GC  
located about 13~kpc from the Galactic center (Harris 1996). Since it is not 
affected by high reddening---$E(\bv) = 0.16$~mag, according to the Harris 
catalogue---we find it somewhat surprising that the {\em only} CMD for this 
cluster available in the literature seems to be the photographic one provided 
by Harris (1975). This is even more puzzling in view of the fact that it has been 
over a quarter of a century since Harris called attention to some intriguing 
characteristics of this cluster's CMD---such as an abnormally high ratio $R$ of 
HB to red giant branch (RGB) stars and the presence of an unusually strong blue 
HB component for such 
a moderately metal-rich cluster. Interest in the CMD of this particular cluster 
seems to have been revived only very 
recently, when Catelan et al. (1998a) pointed out 
that  ``new CCD investigations are especially encouraged [for M75], since the 
latest CMD available (Harris 1975) is strongly suggestive of an NGC~1851-like HB 
bimodality." Note that, due to M75's distance, the Harris photographic CMD was 
only barely able to reach the HB level of the cluster. 

It is the main purpose of the present paper to present the first deep CCD photometry 
of M75, and to investigate whether---as suggested by Catelan et al. (1998a)---M75 
can be classified as yet another bona fide bimodal-HB cluster. In \S2, we discuss 
our observing runs and data reduction techniques. In \S3, the cluster CMD is 
discussed in detail, revealing not only a clearly bimodal HB but also a third 
grouping of blue tail stars clearly separated from the bulk of the blue HB, 
and likely indicating the cooler end of a very extended, though thinly populated, 
blue tail. The presence 
and radial distribution of blue straggler stars (BSS)
is also discussed, and Harris's 
(1975) suggestion of an abnormally high $R$ ratio is addressed. In \S4, we 
evaluate the  ``photometric metallicity" and reddening of the cluster, and in 
\S5 we derive its age relative to NGC~1851. 

In a companion paper (Corwin et al. 2001, hereinafter Paper~II), we will present 
a detailed analysis of the variable star population in the cluster.

\begin{figure*}[t]
      \resizebox{\hsize}{!}{\includegraphics{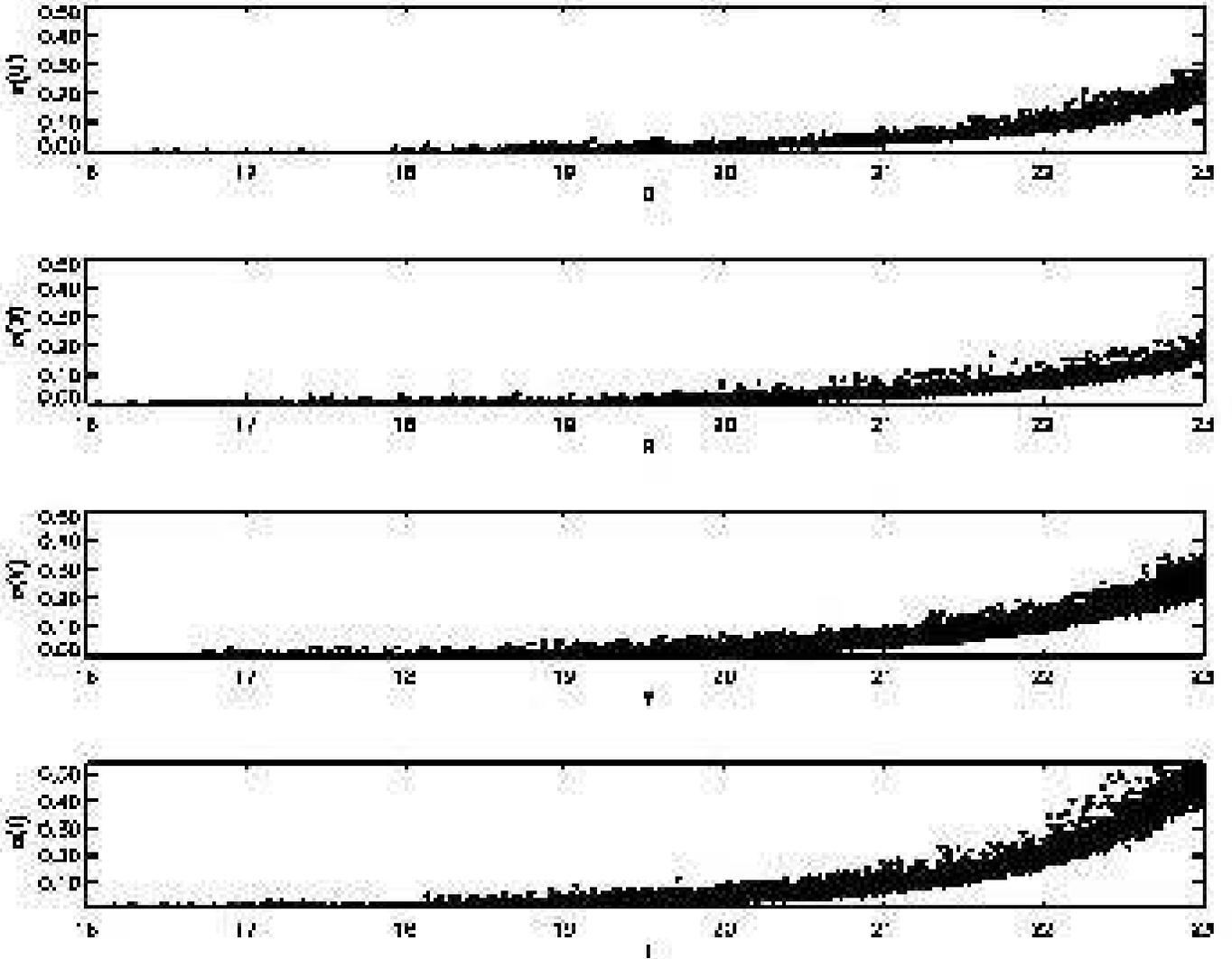}}
      \caption{Formal errors of the NTT photometry from 
      the {\sc daophot} package as a function
      of the mean $U,B,V,I$ magnitudes for each star.
      }
      \label{Fig01}
\end{figure*}

\section{Observations and Data Reduction}

Our analysis is based on the following observational material:

      1. A  set of  ``blue"  $U,B$ and  ``red" $V,I$ CCD frames obtained in
June  1997 on the ESO-NTT  telescope with the $1024\times1024$ and $2048\times1900$
CCD cameras. The array scales were
$0.361\arcsec\,{\rm pixel}^{-1}$ and $0.266\arcsec$ ${\rm pixel}^{-1}$, giving  
fields of view of
about $6.16\arcmin \times 6.16\arcmin$ and $9.08\arcmin \times 8.42\arcmin$. 
The data consist of one short (60-20 sec) exposure image in $B,V,I$, and 
two long exposure images (900-300 sec) in $U,B,V,I$.

      2. Fourteen $B,V$ frames obtained in July  1999 
at the 0.9-meter telescope of the Cerro Tololo Inter-American 
Observatory with a $2048\times 2048$ CCD camera.  The pixel scale of 
 $0.396\arcsec\,{\rm pixel}^{-1}$ gave an observing area 
of $13.5\arcmin \times\ 13.5\arcmin$.

      3. Ten $B,V$ frames obtained in July 1989 
at the 1.54-meter  Danish  telescope  with a $512\times 320$ CCD camera.  The pixel 
scale of 
 $0.47\arcsec\,{\rm pixel}^{-1}$ gave an observing area 
of $4\arcmin \times\ 2.5\arcmin$.
 
The stellar photometry of NTT images was carried out 
separately for all  frames using  {\sc daophot/allstar} (Stetson 1993).
The instrumental magnitudes are transformed photometrically to the deepest reference
frame in each filter. The magnitudes in the same bands of the common unsaturated 
stars 
are averaged.
The brightest stars are measured only on the short exposure time images.  
The field of view of  ``blue" ($U$ and $B$) images is smaller than for  ``red" 
($V$ and $I$) images.
The datasets were  matched together and a final catalog with stellar coordinates 
and 
the instrumental magnitudes in each filter has been compiled for all objects 
identified 
in each field. 
The instrumental values of all the NTT frames were
transformed to the standard system using 20 standard stars 
in five selected fields from Landolt (1992), covering a color 
range $-0.24 < (\bv) < 1.13$. 
The mean photometric errors are $\approx 0.04$~mag for $U,\,B,\,V,\,I < 22$~mag  
and 
between 0.06~mag and 0.1~mag for the fainter magnitudes.
The formal errors for all stars vs. their magnitudes are displayed in Fig.~\ref{Fig01}.
The artificial star 
technique (Stetson \& Harris 1988; Stetson 1991a, 1991b) was used to determine the 
completeness
limits of the data. The $50\%$ completeness limit of the photometry is determined at 
$(U, B, V, I) = (22, 22.4, 22.4, 21.8)$~mag at a distance of $0.8\arcmin$ 
from  the cluster center.

{\sc daophot/allframe} (Stetson 1993) was applied to 28 of the CTIO 
images (14 $B$ and 14 $V$).  All frames are calibrated independently using 13 
photoelectric standards from Alvarado et al. (1990) and Harris (1975).
The photoelectric standards included stars as red as the
reddest M75 red giant stars. 
A  file with the coordinates of 7504 stars and 28 columns of standard magnitudes 
(14 columns of $V$ and 14 of $\bv$) was obtained.  For each star the mean 
$V$ magnitude
and $\bv$ color as well as the standard deviations were determined. 
The mean photometric errors are $\approx 0.08$~mag for $B,\,V < 19$~mag  and
0.12~mag for the fainter magnitudes.
The $50 \%$ limit of the photometry is at 
$(B, V) = (21, 21.2)$~mag at radius $1\arcmin$.

The stellar photometry of 1.54m Danish images was also carried out 
with  {\sc daophot/allstar}. The mean photometric errors are $\approx 0.03$~mag for 
$B,\,V < 19$~mag  and 0.08~mag for the fainter magnitudes.
The $50 \%$ limit of the photometry is at $(B, V) = (18.5, 18.3)$~mag at radius 
$0.3\arcmin$. The seeing for the Danish observing run was better 
($0.9\arcsec$) than for the NTT observing run ($1.2\arcsec$).

 We  transformed the CTIO and Danish $x, y$ coordinate systems to 
the NTT  local system and then searched for stars in common between the two 
datasets. We checked the residuals in magnitude and color for the stars in 
common. They do not show any systematic difference or trend so 
we can conclude that all datasets are homogeneous in magnitude within the errors. 
 
Taking into account the above errors and completeness determination,  
we used the NTT dataset for the
global analysis of the CMD. For star counts we added 28 bright stars  from the 
Danish dataset which are saturated on NTT images and  have radial distances 
between 
$0.3\arcmin$ and $0.8\arcmin$. It should be noted that these 28 stars  
were used {\em only} for number counts. The fiducial line of M75 (see \S3.1 
below) was determined using solely the NTT dataset.

\begin{figure*}[t]
 \centerline{\psfig{figure=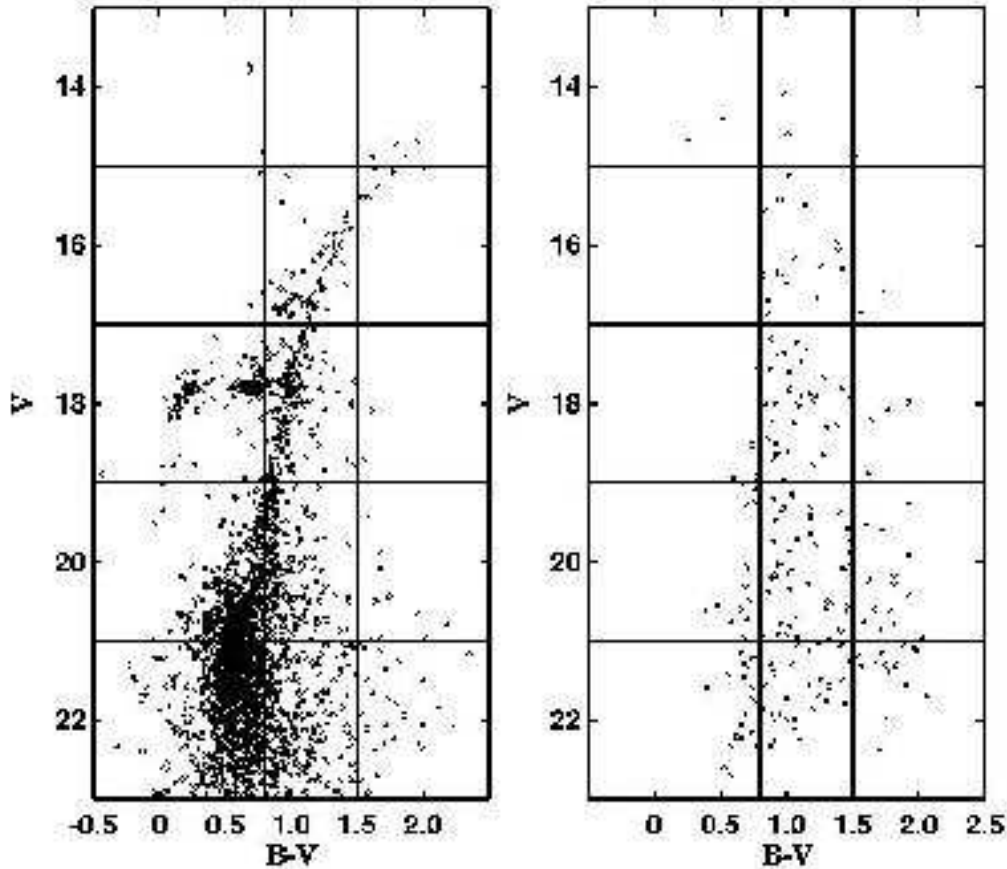}}
 \caption{$(\bv,V)$ color-magnitude diagram for M75 (left panel).
          The field CMD (right panel) is based on images taken 
          $30 \arcmin$ east of the center of the cluster. 
   }
      \label{Fig02}
\end{figure*}

The field stars are statistically decontaminated using off-field CTIO images 
taken  $30\arcmin$ east from the center of the cluster and 
normalized to the same area as the NTT
field of view.  
The $(\bv,V)$ CMDs of ``cluster + field"  and ``field" are  divided
into 15 boxes (shown in
Fig.~\ref{Fig02})---five intervals in magnitude and three in color. 
The stars in each box in the two diagrams are counted. Then, an
equivalent number of stars is removed from the single boxes of the 
``cluster + field" CMD on the basis of the number of field stars found in the
``field" CMD alone (right-hand-side panel in Fig.~\ref{Fig02}).

%

%
\begin{figure*}[t]
  \psfig{figure=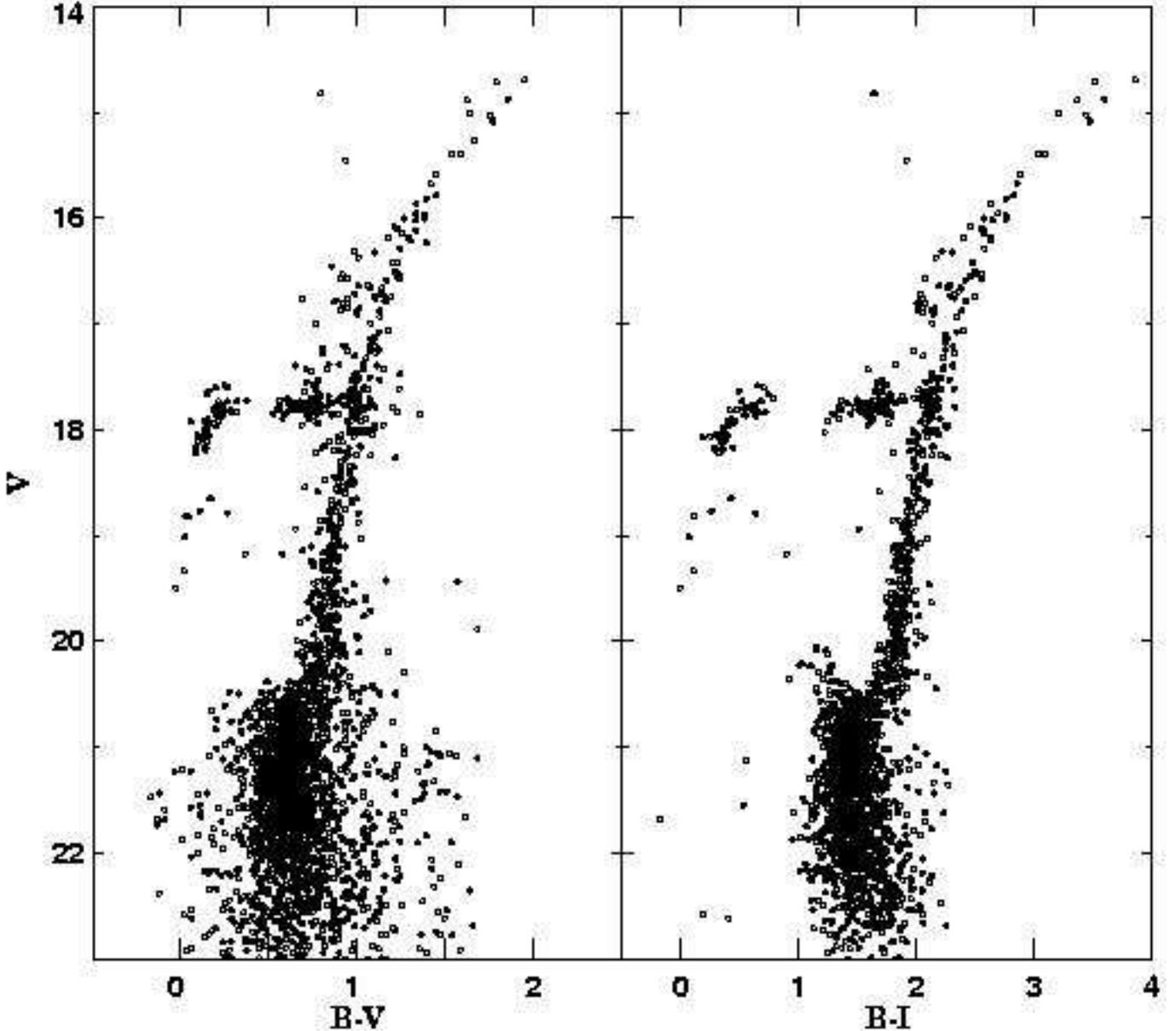}
   \caption{($V$, $\bv$) and ($V$, $B-I$) CMDs  for all stars 
       in the NTT dataset for M75.  
       Known variable stars are omitted from the plot. 
          }
      \label{Fig03}
\end{figure*}

%
\begin{figure*}[t]
	   \centerline{\psfig{figure=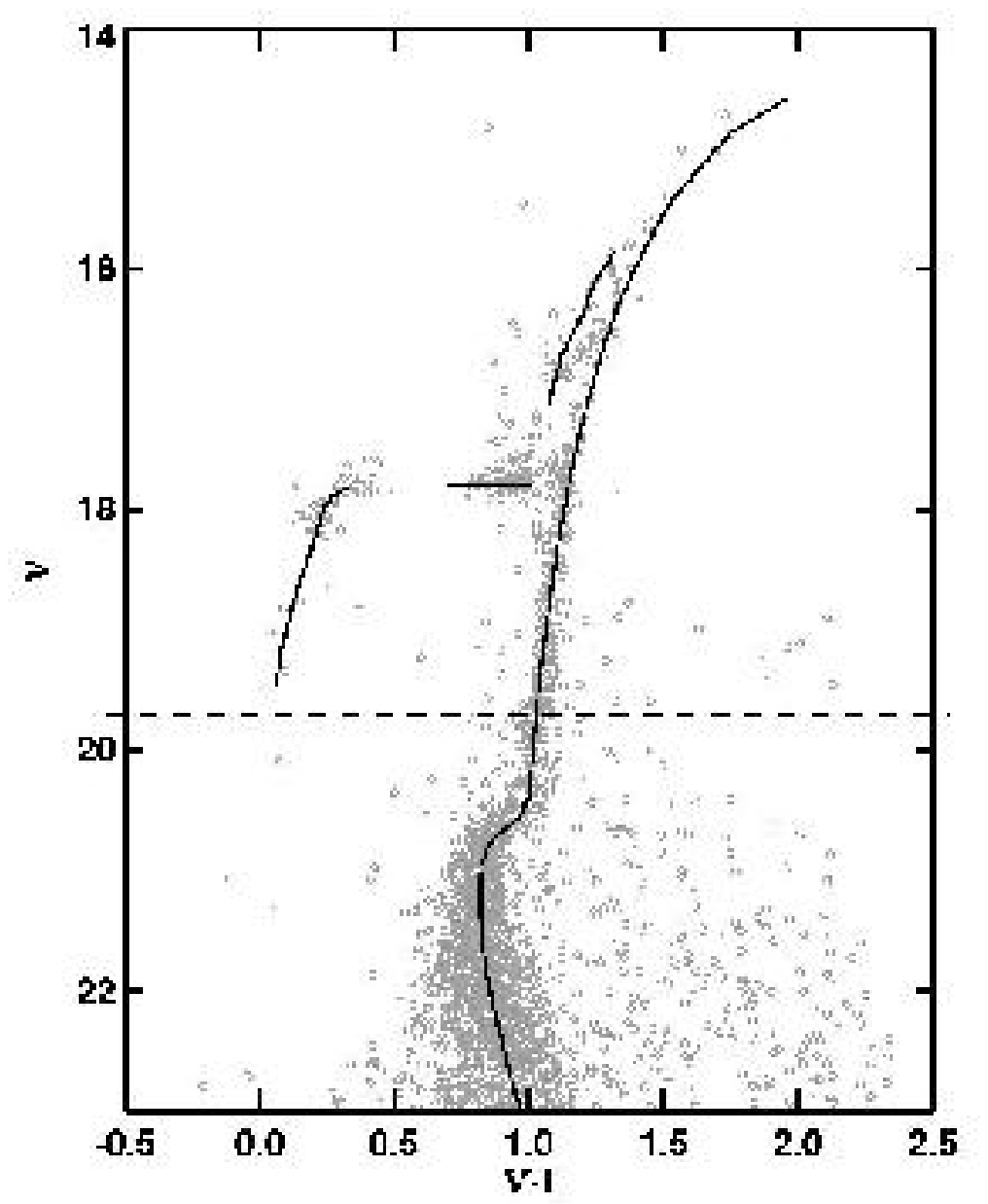}}
      \caption{($V$, $V-I$) CMD for M75, with the derived ridgeline overplotted.  
      Crosses indicate variable stars (from Paper~II). The dashed line indicates 
      the magnitude level which separates the two samples used for  
      better definition of the CMD branches and determination of the ridgeline: 
      while for $V < 19.7$~mag all stars were utilized (and are displayed), 
      fainter than this level  
      only stars with $r > 150\arcsec$ were employed (and are shown). 
      Note that, for the red HB, the ``ridgeline" actually 
      represents the lower envelope of the distribution.
      }
      \label{Fig04}
\end{figure*}

\section{The Color-Magnitude Diagram}

\subsection{Overall CMD Morphology}

The ($V$, $\bv$) and ($V$, $B-I$)  CMDs   from 
the NTT dataset are presented in Fig.~\ref{Fig03}.
Field stars were statistically subtracted as described in \S2,  
and known variables are omitted from the plot.

Mean ridgelines of the main branches of M75 were determined from the 
($V$, $V-I$) CMD. With this purpose, the following 
selection criteria were adopted in order to have the best 
definition of the CMD branches: for $V < 19.7$~mag, all stars were used; 
for $V > 19.7$~mag, and in order to better outline the turnoff (TO) 
region, only stars  with $r > 150\arcsec$  were 
employed (thus avoiding the crowding in the innermost regions 
while minimizing field star contamination).  

The mean ridgelines of the main branches in the CMD  were determined following 
the polynomial fitting
technique by Sarajedini (1994). A first
rough selection of the candidate RGB stars was performed by eye,
removing the HB and part of the asymptotic giant branch 
(AGB) stars. A polynomial law in the
form $(V-I)=F(V)$ was used. In several iterations the
stars deviating by more than 2-sigma in color from the fitting 
relation were rejected. 
The standard error of the 
resulting fit is 0.04~mag.  The mean ridgelines of the fainter 
part of the CMD  ($V > 19.7$~mag---subgiant branch and
main sequence) were determined by 
dividing these branches into bins and computing in each 
bin the mode of the distribution in color. The ``ridgeline" for the 
red HB region was determined
as the lower envelope of the red clump.
Table~1 presents the adopted normal points for each branch. 
In Fig.~\ref{Fig04} the ($V$, $V-I$) CMD is shown with 
the mean ridgelines overplotted.

The main sequence (MS) TO point is found to be at 
$V_{\rm TO} = 21.22\pm0.09$~mag and 
$(V-I)_{\rm TO} = 0.818\pm0.05$~mag.
 The magnitude level of the HB, determined as the lower 
boundary of the red HB  ``clump," is 
$V_{\rm HB} = 17.80\pm0.03$~mag. This implies a magnitude
difference between the HB and the TO of
$\Delta\,V^{\rm HB}_{\rm TO} = 3.42\pm0.09$~mag.

\begin{figure*}[t]
  \centerline{\psfig{figure=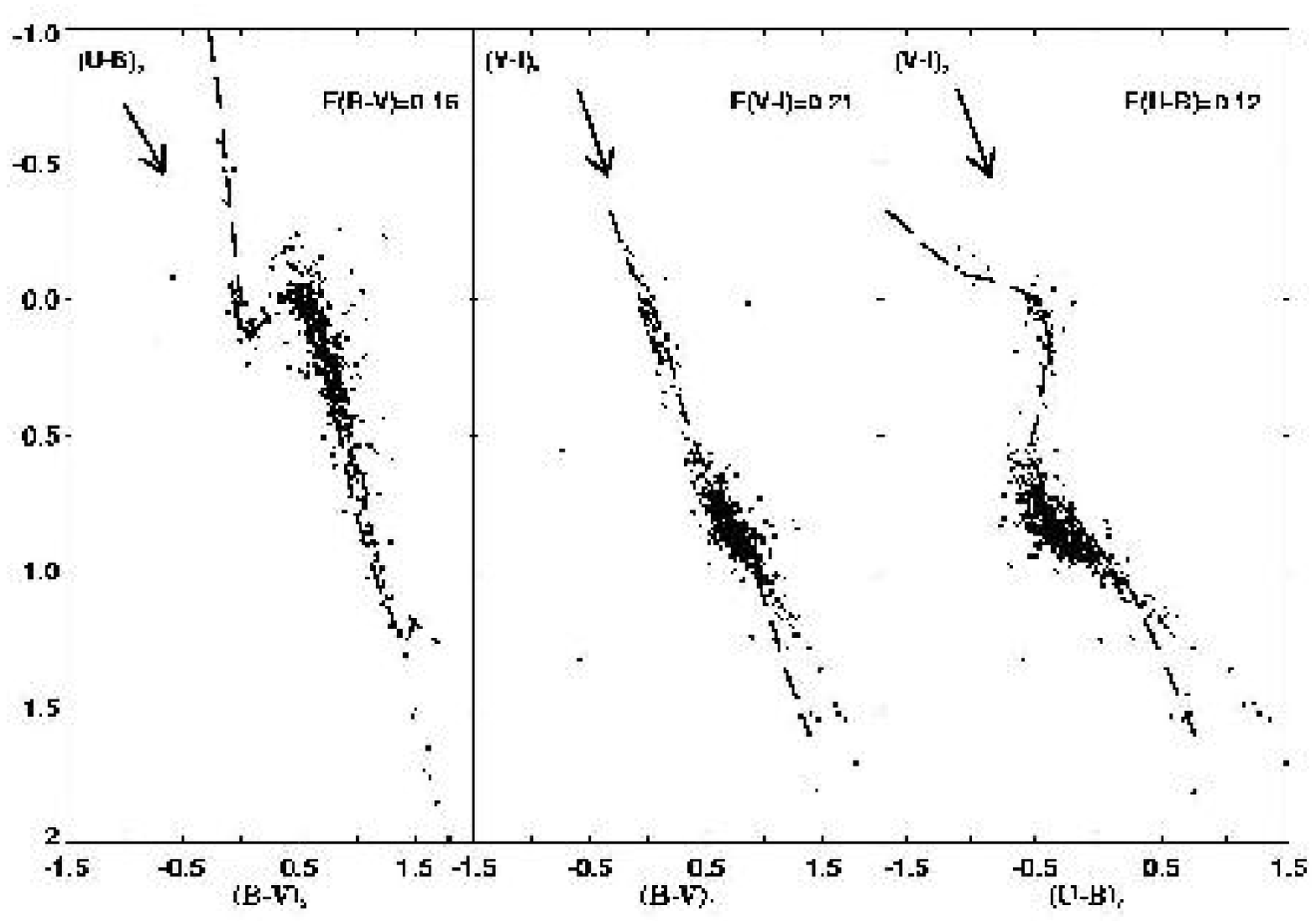}}
  \caption{Dereddened color-color $(\bv)_{0}$ vs. $(U-B)_{0}$, 
  $(\bv)_{0}$ vs. $(V-I)_{0}$, and $(U-B)_{0}$ vs. $(V-I)_{0}$ 
     diagrams. The dashed lines are sequences 
     of luminosity class V from Bessel (1990). The arrows show
     the direction of the reddening vector.  
      }
      \label{Fig05}
\end{figure*}

\subsection{Metallicity and Reddening}

The best fit of 
$(\bv)$ vs. $(U-B)$, 
$(\bv)$ vs. $(V-I)$, and 
$(U-B)$ vs. $(V-I)$ distributions for the M75 blue HB 
stars to the 
color-color lines from Bessel  (1990) gives $E(\bv)=0.16\pm 0.02$. 
The fit is shown on Fig.~\ref{Fig05}.   

For comparison, 
the Schlegel, Finkbeiner, \& Davis (1998) {\sc cobe/dirbe} dust maps give 
$E(\bv) = 0.152$~mag at the M75 position, and Harris (1996) lists 
$E(\bv) = 0.16$~mag in the latest (June~1999) version of his catalogue.  

A complete set of RGB parameters and metallicity  indicators in the
classical $(V,\,\bv)$  plane has been recently presented by
Ferraro et al. (1999a, hereafter F99).   F99 also
reported on an independent calibration of RGB parameters in terms of
the cluster metallicity (both in the Carretta \& Gratton 1997, hereafter 
CG97, [Fe/H] scale and in the corresponding {\it global} [M/H] scale).
In particular, F99 presented a system of equations (see Table~4 in F99)
which can be used to simultaneously derive an estimate of metal
abundance (in terms of ${\rm [Fe/H]}_{\rm CG97}$ and [M/H]) and reddening from
the morphology and location of the RGB in the $(V,\,\bv)$ CMD. 
The ``global metallicity" [M/H] assumes an $\alpha$-element enhancement 
of $[\alpha/{\rm Fe}] = 0.28$ for ${\rm [Fe/H]} < -1$. 

For M75, we used the mean ridgeline  to measure
the following RGB parameters:  
$(\bv)_{\rm 0,g}$, defined as the RGB color at
the HB level; the two RGB slopes $S_{2.0}$ and $S_{2.5}$, defined as the
slope of the line connecting the intersection of the RGB and HB with
the points along the RGB located, respectively, 2.0 and 2.5~mag
brighter than the HB; 
$\Delta V_{1.1}$, $\Delta V_{1.2}$ and $\Delta V_{1.4}$---defined as 
magnitude differences between the HB and RGB at the fixed colors 
$(\bv)_{0} = 1.1, 1.2 $ and $1.4$ mag. 
The derived metallicity, based on 
the weighted mean of the above defined values as listed in Table~2, is 
${\rm [Fe/H]}_{\rm CG97}=-1.01\pm0.12$, ${\rm [M/H]}=-0.82\pm0.19$ 
and  ${\rm [Fe/H]}_{\rm ZW84}=-1.22\pm0.21$.
The derived reddening is $E(\bv)=0.155\pm0.032$. 
The errors associated with the measurements  are conservative estimates of 
the global uncertainties, formal errors being much smaller than those
assumed.

\subsection {Red Giant Branch}

\begin{figure*}[t]
   \centerline{\psfig{figure=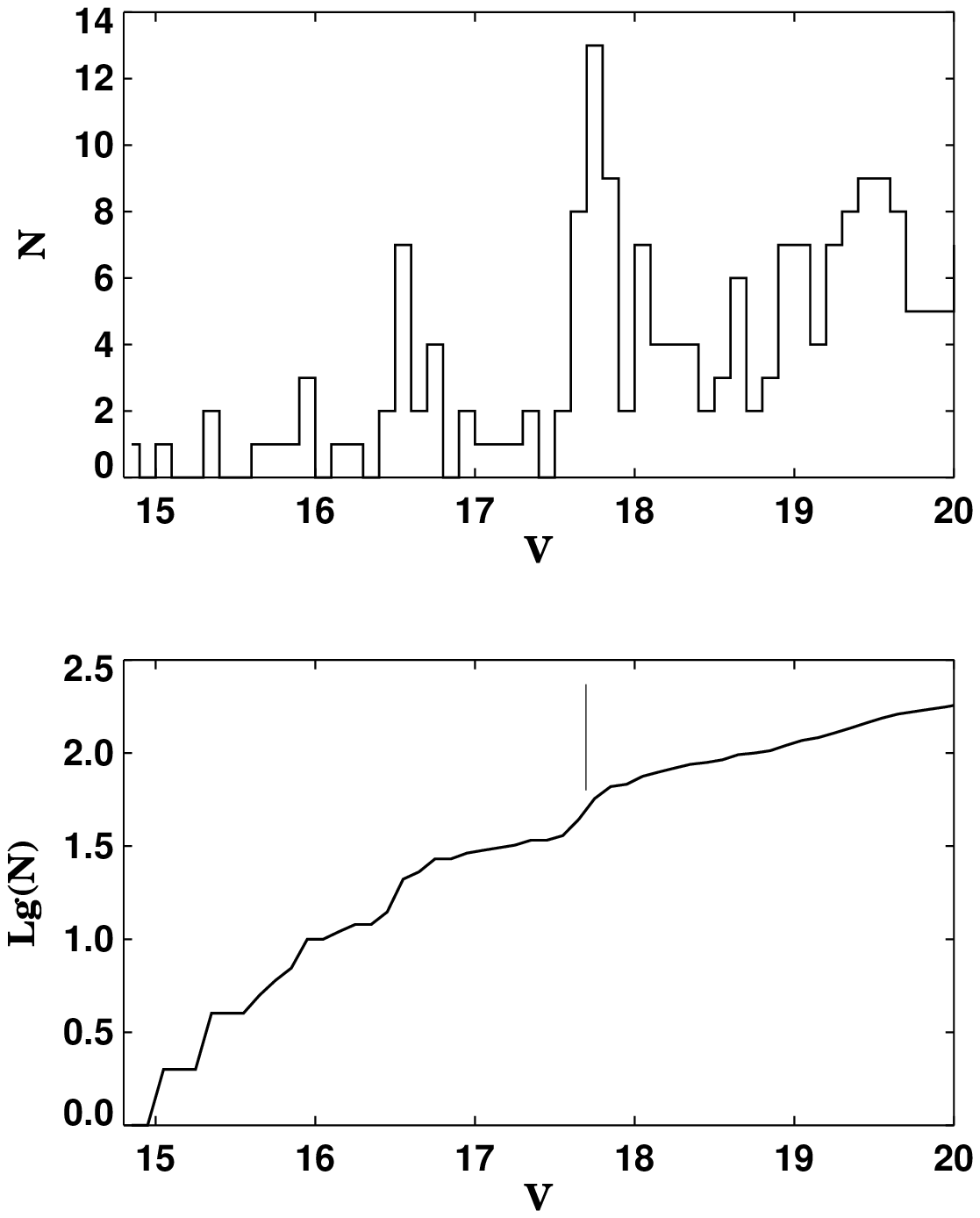}}
   \caption{The observed differential  (upper panel) and integrated 
   (lower panel) luminosity function for the M75 RGB. 
}
      \label{Fig06}
\end{figure*}

The so-called RGB ``bump," corresponding to the point in the 
evolution of RGB stars when the H-burning shell encounters the chemical 
composition discontinuity left behind by the maximum inward penetration 
of the convective envelope, was first predicted by theoretical models 
(Thomas 1967; Iben 1968) and subsequently observed in many clusters (see 
F99 for more details).  
The break in the slope of 
the cumulative RGB luminosity function is a common technique 
used to locate the RGB bump  (Fusi Pecci et al. 1990).
The observed differential and integrated luminosity functions 
for the M75 RGB are shown in Fig.~\ref{Fig06}.
As can be seen the RGB bump of M75  is very well 
defined  at $V=17.75\pm0.05$~mag.

\begin{figure*}[t]
   \centerline{\psfig{figure=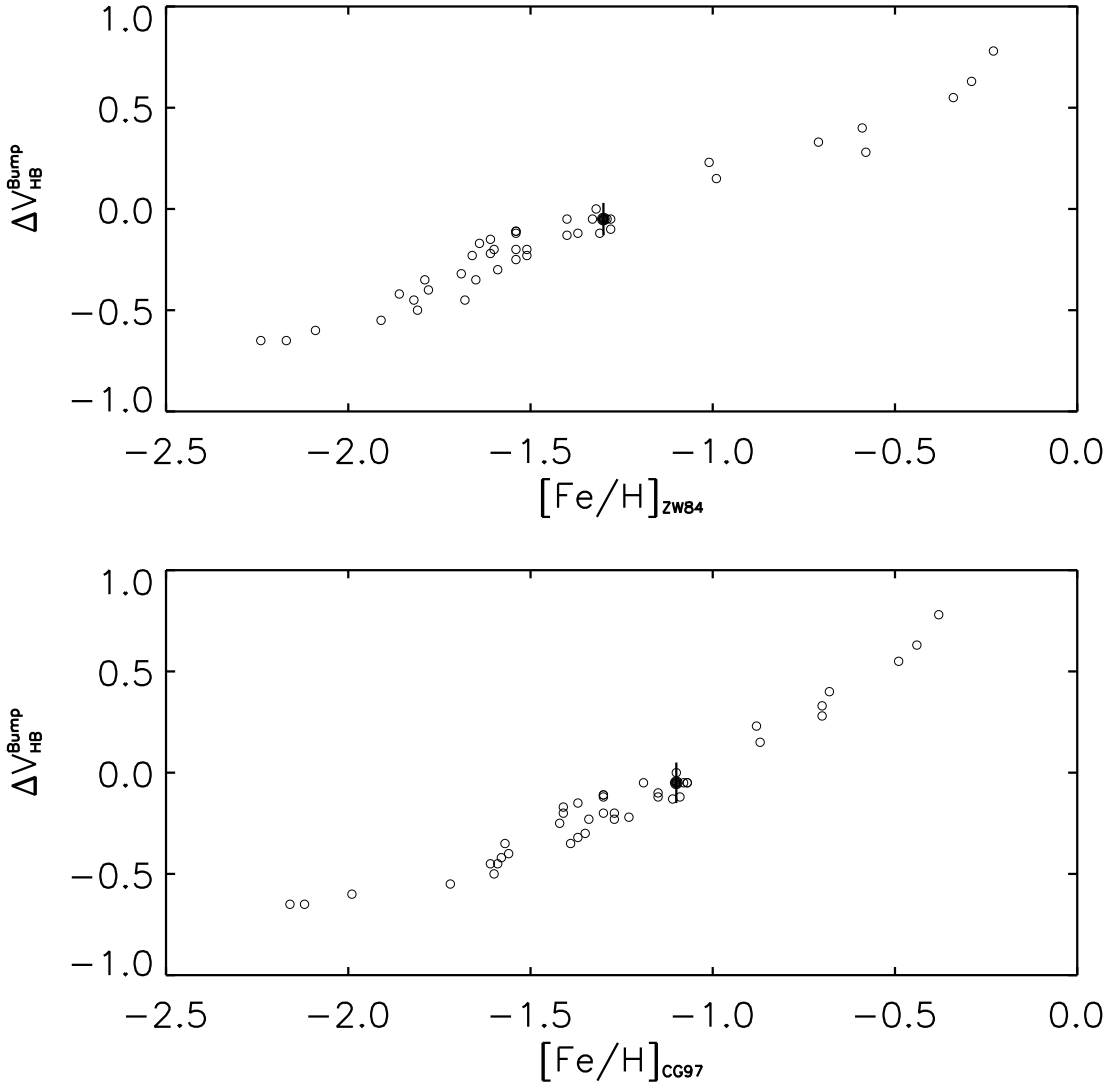}}
   \caption{The magnitude difference between the bump and the HB, 
   $\Delta V_{\rm HB}^{\rm bump}$, as a function of [Fe/H]. 
   Upper panel: CG97 scale; bottom panel: ZW84 scale. The filled circle 
   indicates M75. The clusters data are from Table~5 in F99. 
   }
      \label{Fig07}
\end{figure*}

The magnitude difference between the
RGB luminosity function ``bump" and the HB level, 
$\Delta V_{\rm HB}^{\rm bump}$, is a metallicity indicator recently 
recalibrated by F99. Using the data 
given in their Table~5 and the calibration equations from Table~6 in F99 we 
calculated 
${\rm [Fe/H]}_{\rm CG97}=-1.10\pm0.10$, ${\rm [M/H]}=-0.91\pm0.09$ and 
${\rm [Fe/H]}_{\rm ZW84}=-1.30\pm0.08$. As can be seen, the RGB bump-based 
metallicity is in good agreement with the value derived in \S3.2. 
A plot showing the difference in brightness between the RGB bump and 
the HB, 
$\Delta V_{\rm HB}^{\rm bump}$, as a function of metallicity
is presented in  Fig.~\ref{Fig07}.

Taking the RGB bump-based metallicity into account, 
the revised metallicity of M75, based on the weighted mean 
over {\em all} the values listed in Table~2, 
is ${\rm [Fe/H]}_{\rm CG97}=-1.03\pm0.17$ and ${\rm [M/H]}=-0.87\pm0.19$ 
on the Carretta \& Gratton (1997) scale. On the
Zinn \& West (1984) scale we find ${\rm [Fe/H]}_{\rm ZW84}=-1.24\pm0.21$.

  %

%

%
\begin{figure*}[t]
   \centerline{\psfig{figure=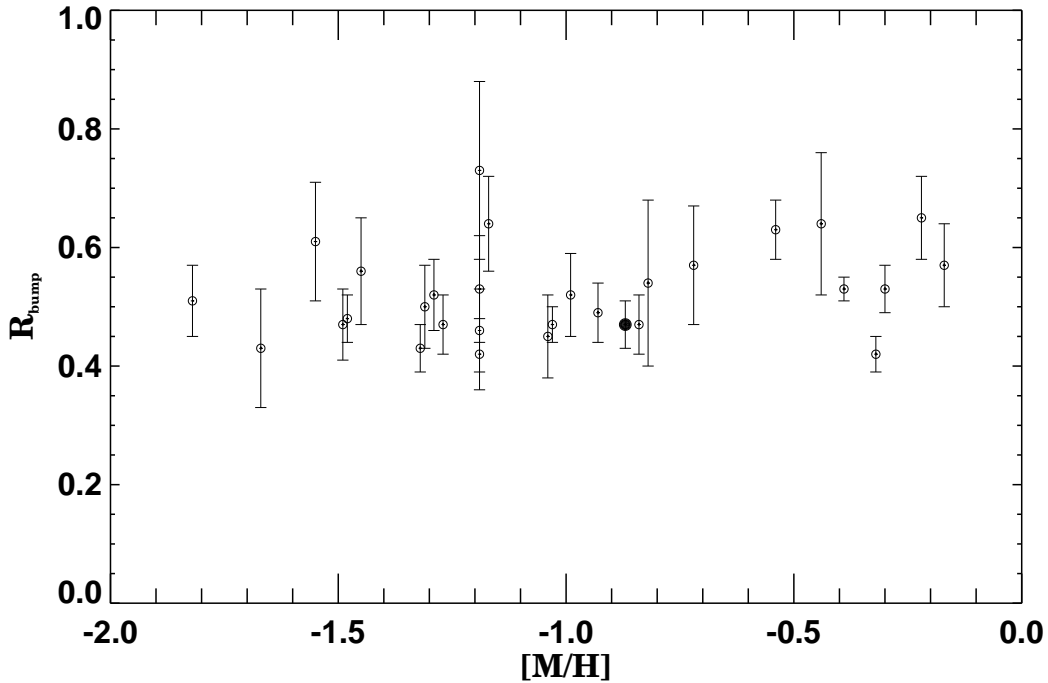}}
   \caption{$R_{\rm bump}$ vs. global metallicity. The filled circle 
      indicates M75.  
   }
      \label{Fig08}
\end{figure*}

%
\begin{figure*}[t]
   \centerline{\psfig{figure=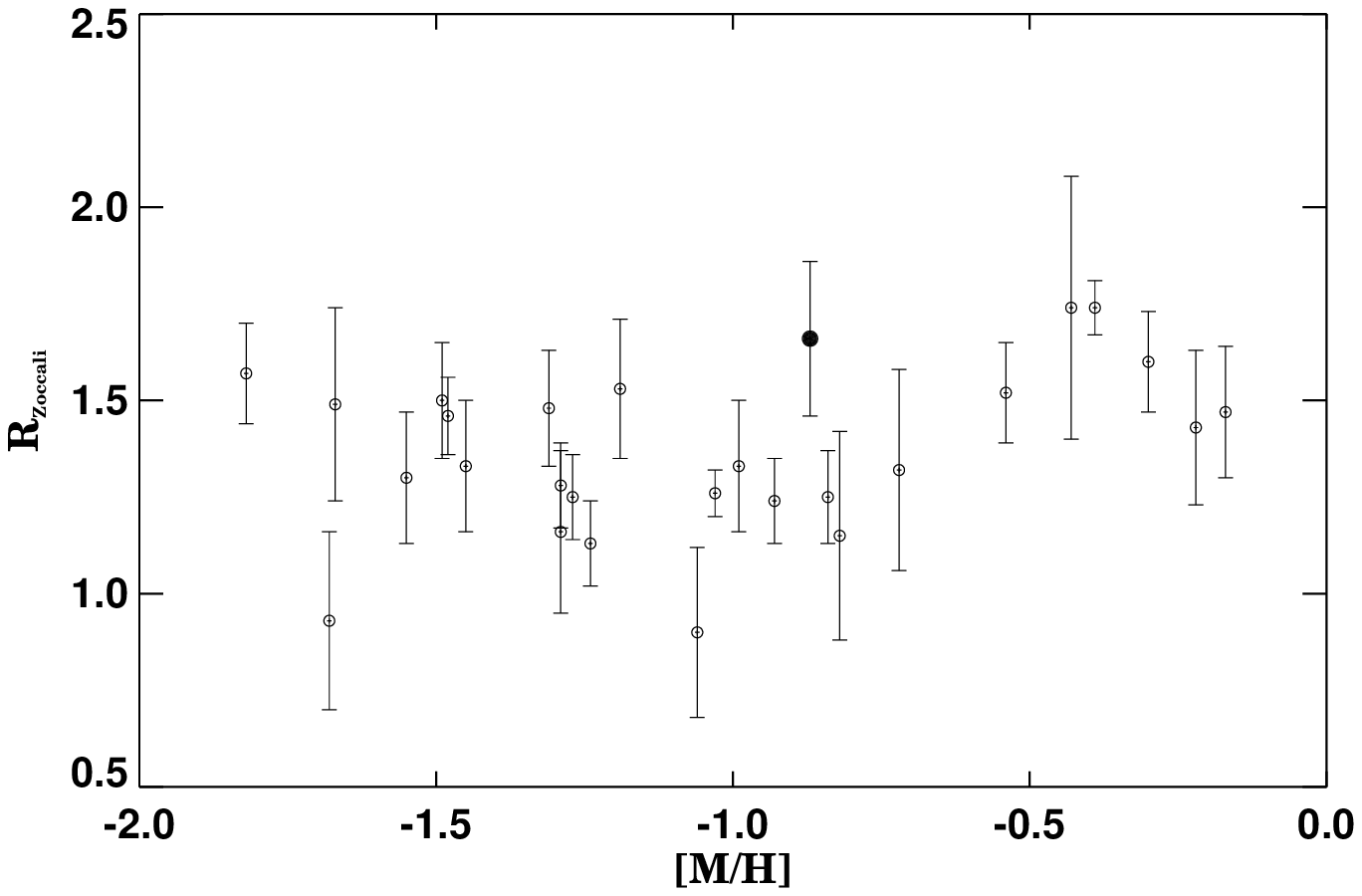}}
   \caption{The $R$-parameter, as defined by Zoccali et al. (2001), is 
   plotted as a function of the ``global metallicity." The filled circle 
   indicates M75. 
   The cluster data are from Table~1 in Zoccali et al. (2001).   }
      \label{Fig09}
\end{figure*}

Recently Bono et al. (2001) defined the new parameter $R_{\rm bump}$ as the 
ratio between the number of stars in the bump region ($V_{\rm bump}\pm0.4$) 
and the number of RGB stars in the interval $V_{\rm bump}+0.5<V<V_{\rm bump}+1.5$.  
In our case we calculated  $R_{\rm bump}=0.47\pm0.04$.

Comparison of the
$R_{\rm bump}$ parameter with the cluster sample of Bono et al. 
(2001)---their  Table~1---shows good agreement. In Fig.~\ref{Fig08}, where 
$R_{\rm bump}$ is plotted as a function of the metallicity, M75 is marked 
with a filled circle, showing its normal location with respect to the 
other clusters. This ``normal" behavior of $R_{\rm bump}$ in M75, in 
analogy with the discussion presented by Bono et al. for the other 
clusters plotted in Fig.~\ref{Fig08}, suggests that the M75 bump too has 
not been affected by the occurrence of non-canonical deep mixing that 
might have substantially altered the chemical composition profile in the 
vicinity of the H-burning shell in the early evolution of M75's red giants.

Harris (1975) first mentioned an abnormally high ratio $R$ of 
HB to red giant branch (RGB) stars. To calculate the $R$ ratio 
we find the following number counts: 
$N_{\rm HB} = 149$,
$N_{\rm RGB} = 118$, and $N_{\rm AGB} = 28$---where $N_{\rm RGB}$ is the 
number of RGB stars brighter than the HB level, following the definition 
by Buzzoni et al. (1983). The observed star counts are 
corrected for completeness and the estimated numbers rounded to the closest 
integer value. It thus follows
that $R=N_{\rm HB}/N_{\rm RGB}=1.26\pm0.15$. For 
$R' = N_{\rm HB}/N_{\rm (RGB+AGB)}$,
the value is $R'=1.02\pm\,0.10$. For
$R1 = N_{\rm AGB}/N_{\rm RGB}$ we calculated 
$R1=0.24\pm\,0.05$ and  $R2 = N_{\rm AGB}/N_{\rm HB}$ is
$R2=0.19\pm\,0.04$. We have assumed, following Buzzoni et al., a 
differential bolometric correction between the HB and the RGB of 0.15~mag. 

If we adopt the Zoccali et al. (2001) definition of  
$N_{\rm RGB}$ as RGB stars brighter than the
ZAHB (which is approximately equivalent to the lower envelope of the
red HB), $N_{\rm RGB}$ becomes 90 (after completeness correction)  and  
$R=1.66\pm0.20$, $R'=1.26\pm\,0.16$ and $R1=0.31\pm\,0.07$. 
Plotting the
$R$ parameter vs. global metallicity in Fig.~\ref{Fig09}  (the cluster 
data are from Table~1 in Zoccali et al.)  we can see that M75 
has a high $R$ value for its metallicity---though only at the $1-2\sigma$
level.  

Note that the $R$ value following the Buzzoni et al. (1983) 
definition does not seem atypical (or large). It is only when we use 
the Zoccali et al. (2001) definition that we are able to verify the 
suggestion by Harris (1975) of a low ratio of giants to HB stars in 
M75. This problem is due to the fact that the bump, at such a metallicity, 
is critically located close to the HB level (see Fig.~7), and may or may 
not be included in the number counts due to only a small change in the 
magnitude cutoff. 

In order to further investigate this issue, we have utilized the observed 
RGB number counts down to a deeper magnitude level on the CMD, performing 
a direct comparison against similarly computed values from the theoretical 
simulations discussed in \S3.4.1 and presented in Figs.~11 and 12 below. In 
order to avoid the ambiguity brought about by the positioning of the RGB 
bump (see Fig.~7) in the derived number counts, we define 
$R_{\langle M_V\rangle+1.0}$ and
$R_{\langle M_V\rangle+1.5}$ as the ratio between HB stars and RGB 
stars brighter than $\langle M_V({\rm HB})\rangle+1$~mag and 
$\langle M_V({\rm HB})\rangle+1.5$~mag, respectively. 
The resulting comparison 
between the observed and theoretical number ratios is given in Table~3. 
As can be seen, the observed $R$-ratios are clearly confirmed to be 
higher than the values predicted for $Y_{\rm MS} = 0.23$, in agreement 
with the conclusion based on the Zoccali et al. (2001) definition of 
the $R$-ratio and with the original suggestion by Harris (1975) that 
M75 seems to have too few giants compared to HB stars.

What is the origin of the larger $R$-ratio for M75? The traditional 
explanation would involve a larger helium abundance for this cluster 
(Iben 1968). Is this a viable explanation, in the case of M75? In 
order to answer this question, we may take benefit of the predicted  
dependence of HB luminosity on the helium abundance, which should 
accordingly lead to longer periods for M75's RR Lyrae variables (e.g., 
Catelan 1996). We 
will check whether this is the case, or whether other explanations 
are required, in Paper~II, where we present a detailed analysis of 
the M75 RR Lyrae light curves.

\subsection{HB Morphology}

\subsubsection{HB Multimodality} 

\begin{figure*}[t]
   \centerline{\psfig{figure=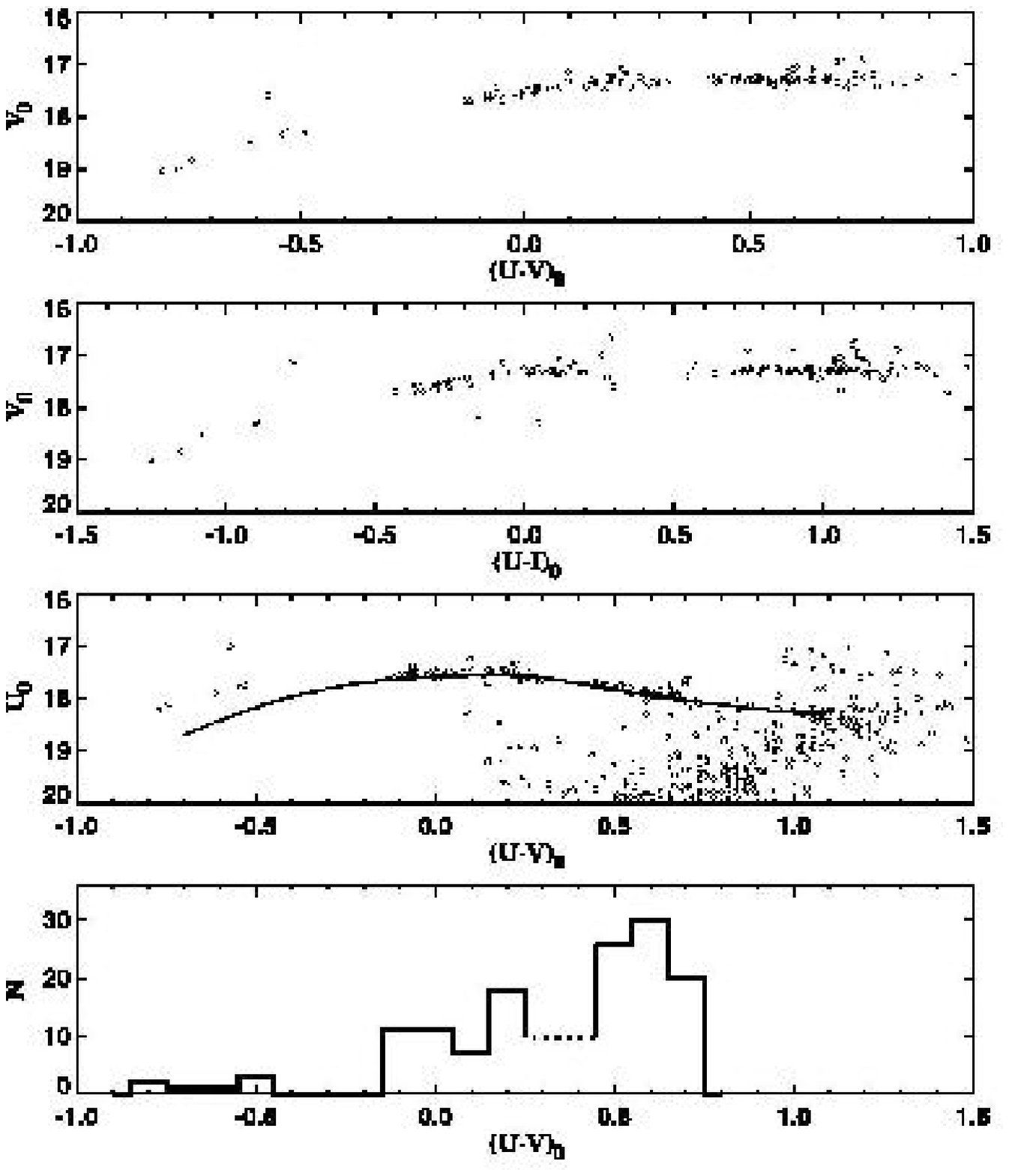}}
   \caption{Zoomed ($V$, $U-V$), ($V$, $U-I$) and ($U$, $U-I$) 
      CMDs of M75 around the HB region. The bottom panel shows a
      histogram of the $(U-V)_{0}$ color distribution for M75 HB stars. 
      In this panel, the RR Lyrae region is indicated with a dashed line; 
      the $(U-V)_0$ color distribution of the M75 RR Lyrae was assumed to 
      be approximately uniform. 
      } 
      \label{Fig10}
\end{figure*}

Fig.~\ref{Fig10}  shows the HB region of the CMD of M75 in several different 
planes: ($V$, $U-V$); ($V$, $U-I$); and ($U$, $U-V$). 
Known variable stars  (see Paper~II) are omitted from these plots. 
Both the red and the blue 
HB regions are very well populated, as first noted by Harris (1975). The blue 
part of the HB has a blue tail and  an obvious  ``gap" at $V \simeq 18.5$~mag 
and $(U-V)_{0}$ between $-0.55$ and $-0.15$, corresponding to temperatures 
in the range $10,\!500 \lesssim T_{\rm eff}({\rm K}) \lesssim 13,\!500$---which 
overlaps the suggested ``G1" and ``G2" gaps in Ferraro et al. (1998). Another, 
much less obvious gap (or underpopulated region) {\em may} be present at 
$V \approx 17.9-18.0$~mag and $(U-V)_0 \approx 0.1$~mag, 
$(B-I)_0 \approx 0.1$~mag.

In order to more quantitatively describe the morphology of the HB, we
computed several useful HB morphology parameters, including those defined 
by Mironov (1972), Zinn (1986), Lee, Demarque, \& Zinn (1990), Buonanno 
(1993), Fusi Pecci et al. (1993), and Catelan et al. (2001b). Several of 
these authors have emphasized the importance of utilizing as many HB 
morphology parameters as possible when drawing general inferences based 
on the HB morphology of Galactic GCs. The values of the computed 
parameters are given in 
Table~\ref{TabHB}. We briefly recall that 
$B$, $V$, and $R$ are the number of blue HB, variable (RR Lyrae), and red 
HB stars, respectively; 
$B2$ is the number of blue HB stars bluer than $(\bv)_0 = -0.02$~mag; 
$B2'$ is the number of HB stars redder than $(V-I)_0 = -0.02$~mag; 
$B7'$ is the number of blue HB stars redder than $(V-I)_0 = +0.07$~mag; 
and, finally, 
$B0'$ is the number of blue HB stars redder than $(V-I)_0 = 0.0$~mag.

To compute the reddening-dependent HB morphology 
parameters, we have assumed 
$E(\bv)=0.16$ (\S3). Also, the number of variable stars, $V = 22$, was 
adopted from our comprehensive variability survey, described in Paper~II. 
All observed number counts are corrected  for completeness.

It is obvious from these numbers that M75 is indeed another cluster 
with a strongly bimodal HB, since 
it has many fewer RR Lyrae variables than either red HB or blue HB stars. 
A histogram with the $(U-V)_0$ color distribution of the M75 
HB stars (corrected  for completeness) is given in the bottom panel of  
Fig.~\ref{Fig10}. 
Since we do not have $U-V$ mean colors for the variables we  assume a 
flat distribution. This clearly shows that the distribution of stars 
along the HB of M75 has three main modes---though the third mode, on 
the blue tail, is very thinly populated. For a detailed discussion of 
the completeness of our RR Lyrae search, see Paper~II. 
 
The M75 HB clearly bears striking resemblance to that of the well-known 
bimodal-HB GC NGC~1851, as first pointed out by Catelan et al. (1998a) 
on the basis of the Harris (1975) photographic CMD. In particular, the 
number ratios $B:\,V:\,R = 0.38:\,0.16:\,0.45$ are  similar to 
those for NGC~1851, for which $B:\,V:\,R = 0.30:\,0.10:\,0.60$. 

We have investigated further the issue of HB bimodality in M75 by 
computing a new 
grid of synthetic HBs aimed at matching the HB morphology parameters 
displayed in Table~4. Our purpose is to check whether a bimodality in 
the underlying evolutionary parameters is necessarily implied by the 
observed bimodality in the HB number counts (Catelan et al. 1998a). Our 
simulations, computed with {\sc sintdelphi} (Catelan et al. 2001b), 
utilize the evolutionary tracks for $Y_{\rm MS} = 0.23$, $Z = 0.002$ 
from Sweigart \& Catelan (1998). Observational scatter has been added 
by means of suitable analytical representations of the errors in the 
photometry. The number of HB stars in these simulations is similar 
to the observed one, $N_{\rm HB} = 149$, but has been allowed to 
fluctuate according to the Poisson distribution.

%
\begin{figure*}[t]
  \centerline{\epsfig{file=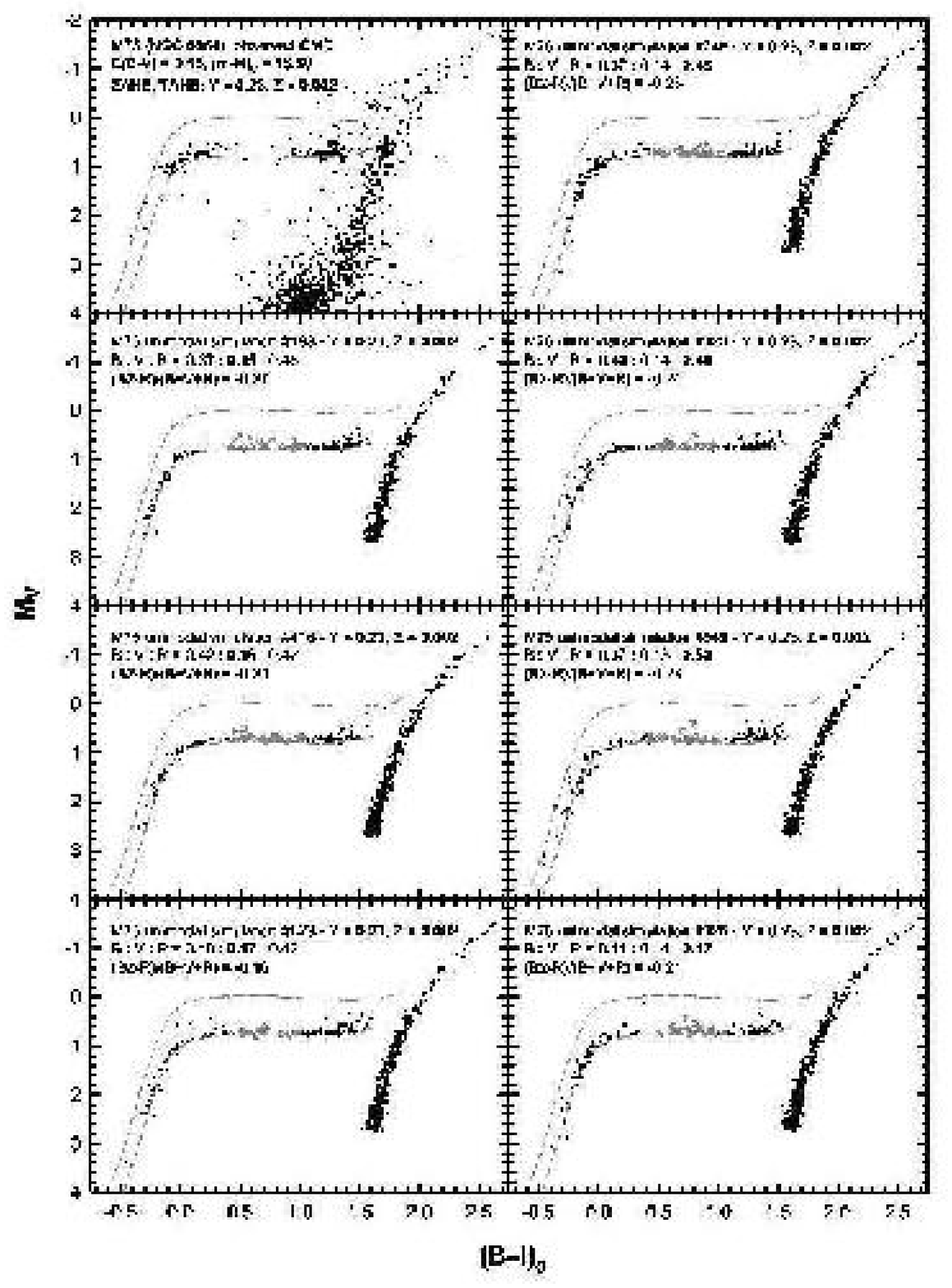,height=6.5in,width=5in}}
  \newpage
  \caption{Random sample of HB simulations for M75 in the {\em unimodal} 
  ZAHB mass distribution case. The $M_V$, $(B-I)_0$ plane is shown. We 
  have assumed $E(B-I)/E(\bv) = 2.7$, after Stetson (1998).  
  In the upper left panel, the observational data are shown, along with 
  the reddening value and distance modulus found more suitable to fit 
  the theoretical ZAHB for the indicated chemical composition to the 
  observed distribution. Variable stars are omitted in the panel
  showing the observed CMD, but not in the others (where they are 
  displayed as encircled gray dots)---which show 
  the theoretical CMD simulations. Also indicated are the $B:V:R$  
  number ratios and the value of the Buonanno (1993) HB morphology 
  parameter. 
    }
      \label{Fig11}
\end{figure*}

%
\begin{figure*}[t]
  \centerline{\epsfig{file=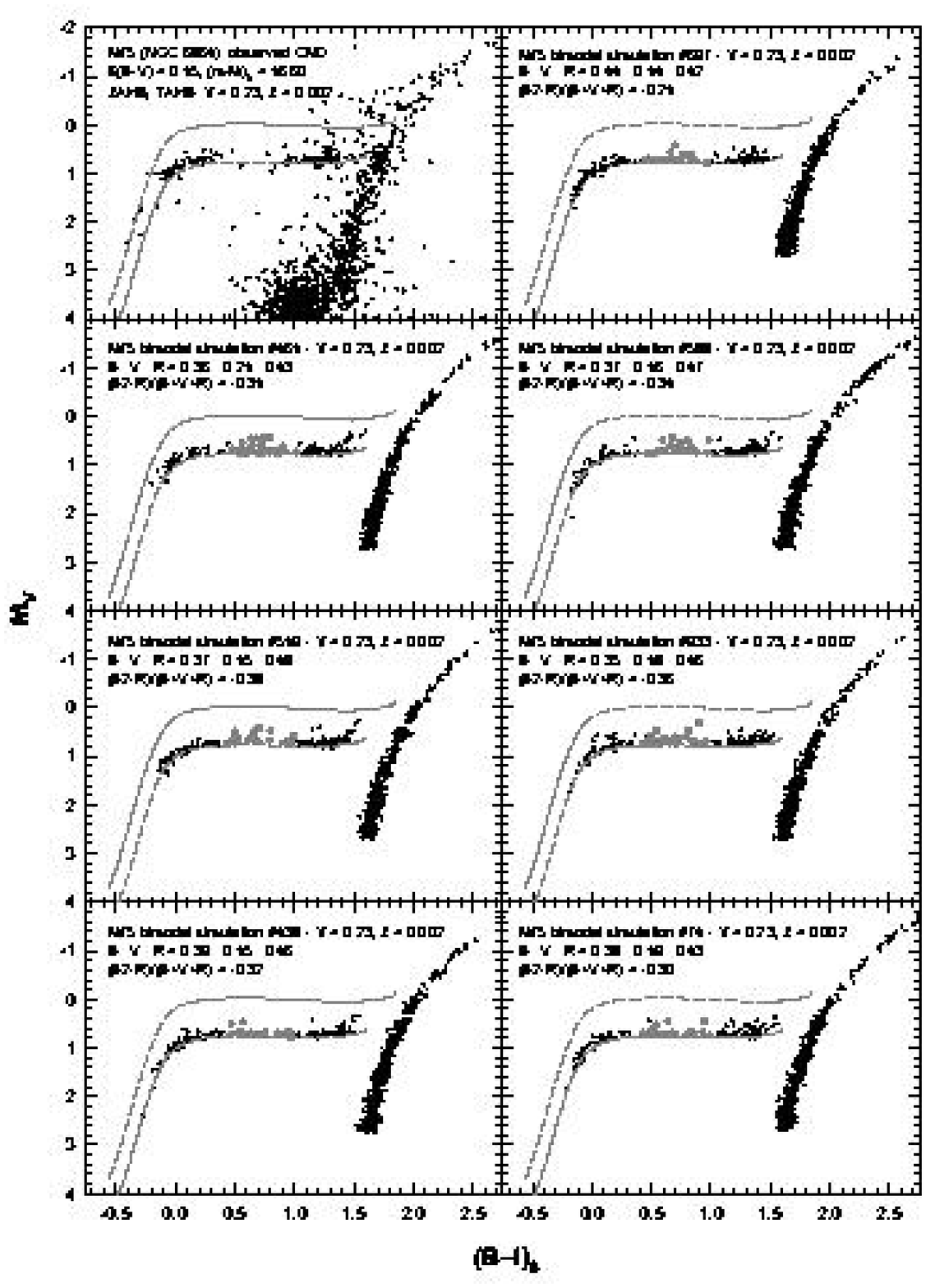,height=7.5in,width=5.5in}}
  \caption{As in Fig.~11, but for the {\em bimodal} 
  mass distribution case.    
    }
      \label{Fig12}
\end{figure*}

We first attempted to reproduce the observed HB morphology parameters 
for M75 by means of a unimodal Gaussian deviate in the zero-age HB 
(ZAHB) mass. Following a procedure similar to that described in 
Catelan et al. (2001a, 2001b), we find a best match between the 
observed and predicted HB morphology parameters for 
$\langle M_{\rm HB}\rangle = 0.624\,M_{\sun}$, 
$\sigma_M = 0.043\,M_{\sun}$. Some randomly selected samples of 
simulations for this unimodal mass distribution case are displayed,  
in the [$M_V$, $(B-I)_0$] plane, in Fig.~11. 
The predicted HB morphology parameters 
for this case are shown in Table~4 (third column); the indicated 
``error bars" represent the standard deviation of the mean over a set 
of 1000 simulations. It is readily apparent that, while the overall 
general agreement, particularly in terms of the commonly employed 
Mironov and Lee-Zinn parameters, is reasonable, some of the other 
parameters introduced by Catelan et al., especially $B0'/B$ and 
$B7'/(B+V+R)$, are not accounted for in an entirely satisfactory way 
by these 
simulations. For this reason, we have attempted to compute extra sets 
of simulations 
for an intrinsically bimodal distribution in mass on the ZAHB.

In the bimodal case, 
we have found that the following combination of parameters gives the 
best fit between the models and the observations: 
$\langle M_{\rm HB,1}\rangle = 0.595\,M_{\sun}$, 
$\sigma_{M,1} = 0.010\,M_{\sun}$, $N_{\rm HB,1} = 46$; 
$\langle M_{\rm HB,2}\rangle = 0.640\,M_{\sun}$, 
$\sigma_{M,2} = 0.030\,M_{\sun}$, $N_{\rm HB,2} = 104$. 
Some randomly selected samples of 
simulations for this bimodal mass distribution case are displayed,  
in the [$M_V$, $(B-I)_0$] plane, in Fig.~12. The predicted HB morphology 
parameters for this case are also shown in Table~4 (fourth column). 

We find that the intrinsically bimodal mass distribution 
case provides a much more satisfactory fit to the whole set 
of HB morphology parameters than does the unimodal case. To some 
extent, this is not an unexpected result, given that this case 
contains five free parameters, as opposed to two in the unimodal 
one (Catelan et al. 1998a). The main 
difference between the bimodal and unimodal cases, as can be readily 
appreciated by inspection of Figs.~11 and 12, is that the unimodal 
case clearly predicts that a fairly extended blue tail should be 
present in M75, whereas this is not necessarily required in the 
bimodal case. Also in this respect, apart from a few cases where 
statistical fluctuations change the predicted simulations quite 
substantially, the bimodal case appears much more successful at 
matching the observed distribution, showing, in general, a more 
stubby blue HB with relatively few hot stars being in general 
expected. Note, however, that neither the unimodal nor the bimodal 
cases shown in Figs.~11 and 12 can successfully account for the 
existence of a third mode of hotter HB stars along the blue tail 
of M75 (see also Fig.~10, bottom panel).

%
\begin{figure*}[t]
   \centerline{\psfig{figure=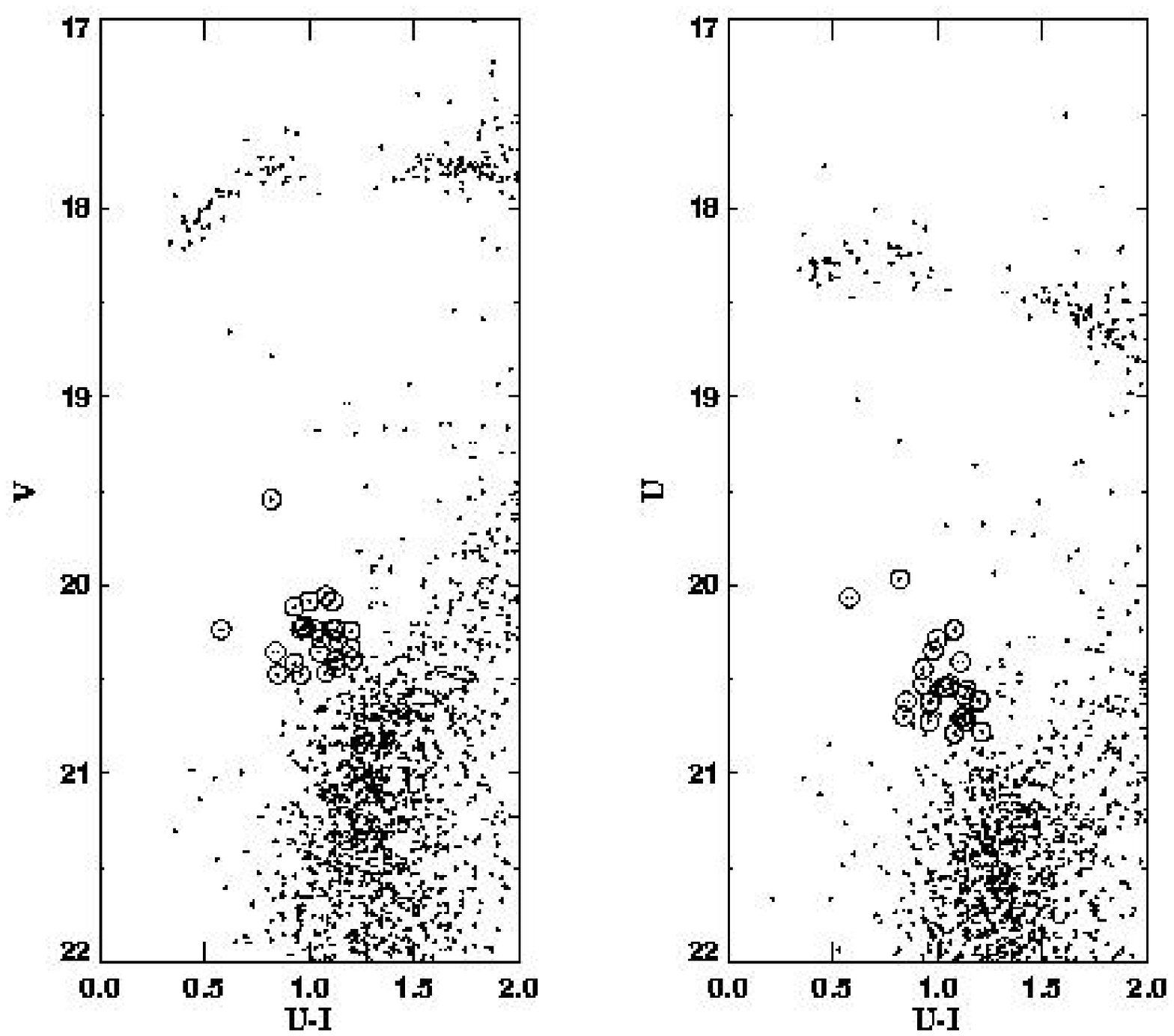}}
   \caption{Zoomed ($U-I$, $V$) and ($U-I$, $U$) CMDs of the BSS region
      (left and right panels, respectively).
      Candidate BSS are plotted as encircled dots.
   }
      \label{Fig13}
\end{figure*}

\begin{figure*}[t]
      \centerline{\psfig{figure=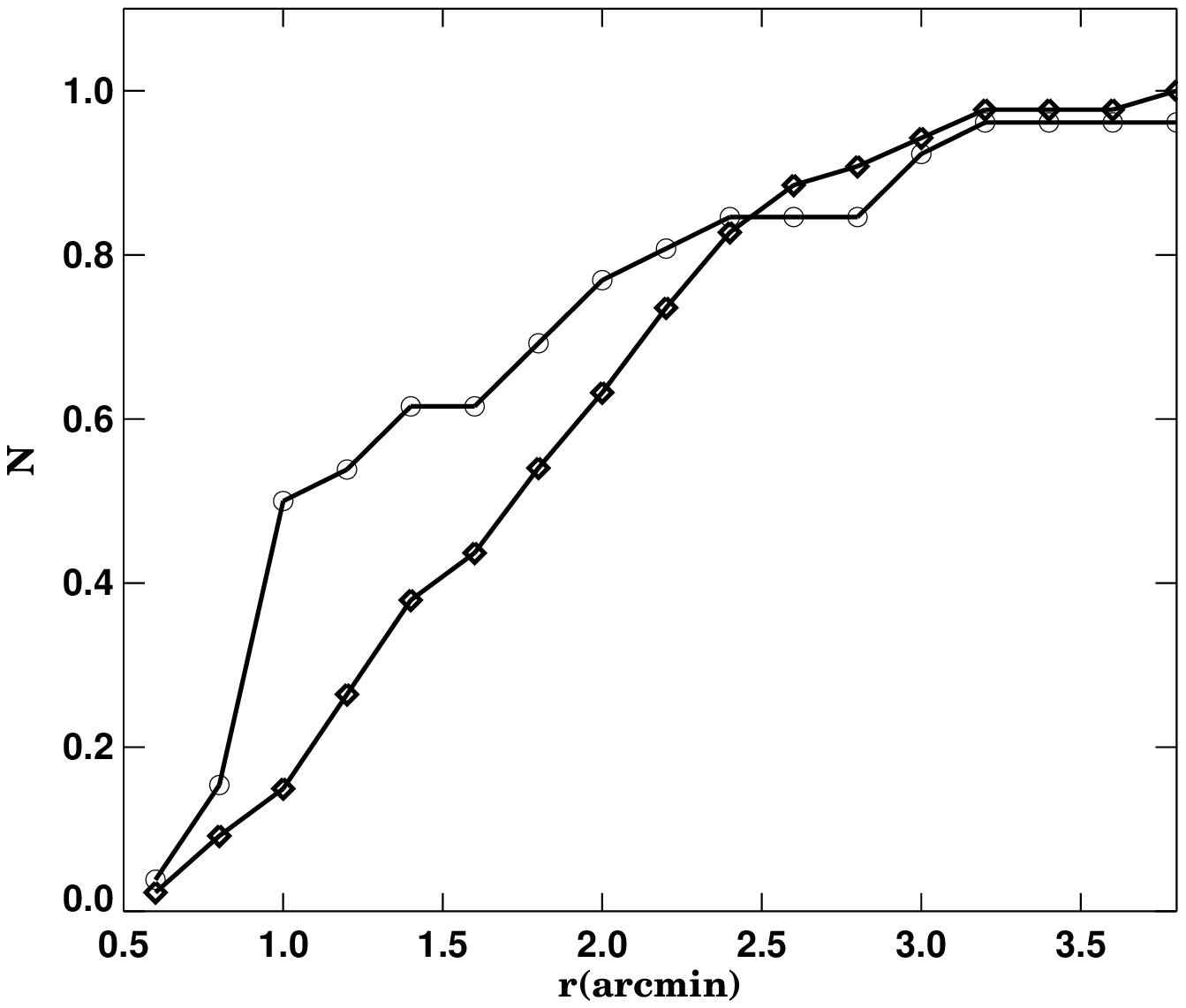}}
      \caption{Radial cumulative distributions of the candidate
      BSS (circles) and SGB stars  (diamonds) in  M75.
      }
      \label{Fig14}
\end{figure*}

\subsubsection{The Grundahl Jump in Broadband $U$} 

Grundahl et al. (1999) have recently shown that the interpretation 
of Str\"omgren $u$ photometry of blue HB stars in GCs is affected 
by a most intriguing phenomenon: at temperatures higher than 
$T_{\rm eff} \approx 11,\!500$~K, all GC CMDs show 
an intriguing ``jump" in the $u$, $(u-y)_0$ plane, with all blue 
HB stars hotter than 11,500~K appearing brighter and/or hotter 
than predicted by the models. As argued by Grundahl et al., and 
spectacularly confirmed by Behr et al. (1999), the reason for this 
discrepancy can be attributed to the fact that the atmospheres of 
hot HB stars in GCs are affected by radiative levitation of metals, 
which pumps up heavy metals to supersolar levels in these stars' 
atmospheres. 

As can be seen in Fig.~10 (bottom panel), a similar jump is present 
also in broadband $U$, with all stars blueward than $(U-V)_0 = -0.5$ 
falling systematically above the canonical ZAHB. The displayed ZAHB 
sequence corresponds to VandenBerg et al. (2000) models for 
${\rm [Fe/H]} = -1.14$ and $[\alpha/{\rm Fe}] = +0.3$, and were 
transformed to the observational plane using standard prescriptions 
from Kurucz (1992)---which assume that the cluster metallicity is 
appropriate also for the atmospheres of hotter blue HB stars. It is 
worth noting that the Grundahl jump phenomenon in broadband $U$ has 
recently been verified also by Siegel et al. (1999), Bedin et al. 
(2000), and Markov, Spassova, \& Baev (2001). The presence of the 
jump phenomenon in broadband $U$ is in good 
agreement with the radiative levitation scenario, as can be seen 
from inspection of Fig.~7 in Grundahl et al. (1999).

\subsection{Blue Straggler Stars}

On the basis of their location in the ($V$, $U-I$) CMD, we identified a 
sample of  26 candidate BSS.  
The $x$ and $y$
coordinates (in pixels)  and  $U$, $B$, $V$, $I$
magnitudes and distance from the cluster 
center (in arcmin) for these candidate BSS are listed in Table~5.

Fig.~\ref{Fig13} shows zoomed M75 CMDs where the BSS candidates are 
plotted as encircled dots. As can be seen, all the candidates selected
in the ($V$, $U-I$) plane lie also in the BSS region in the ($U$, $U-I$)
CMD. Moreover, the latter CMD suggests that a few additional objects 
might be considered BSS candidates. We conservatively decided, however, 
to limit our analysis to those stars which are BSS candidates in both 
the indicated CMDs. 

Careful checks were made to ensure 
that the internal errors of the BSS candidates are the same as 
those of the subgiant branch (SGB) stars at the same brightness
level. The {\sc daophot} 
parameters {\sc sharp} and {\sc chi} were checked for each BSS 
candidate, in order to ensure that most are unlikely to be the 
result of spurious photometric blends (e.g., Ferraro, Fusi Pecci, \& 
Buonanno 1992). 
The positions of the BSS were checked on the statistically 
decontaminated CMDs. 
The number of field stars is not sufficient to explain the  
BSS population (see also Fig.~2).

In order to quantitatively compare the BSS populations among different 
Galactic GCs, one needs to suitably normalize the numbers of such stars. 
In this sense, Ferraro et al. (1999b) defined the specific fraction 
of BSS as the number of BSS normalized to
the number of HB stars observed in the same cluster region, 
$F_{\rm HB}^{\rm BSS}=N_{\rm BSS}/N_{\rm HB}$. In
M75, the specific frequency calculated in this way is 
$F_{\rm HB}^{\rm BSS} \simeq 0.15$.

Generally, BSS are more centrally concentrated with respect to the
other stars in the cluster, although there are notable exceptions to
this general rule, such as M3 (Ferraro et al. 1993, 1997) and M13
(Paltrinieri et al. 1998).  In M75 we  compared the radial
distribution of BSS  with respect to SGB 
stars (in the range
$19.5<V<20.5$). The cumulative radial distributions are plotted, as a
function of the projected distance ($r$) from the cluster center, in
Fig.~\ref{Fig14}, where one sees that the M75 BSS (circles) are more 
centrally concentrated than the SGB sample (diamonds).  The
Kolmogorov-Smirnov test shows that the probability of drawing the two
populations from the same distribution is  $0.3\%$.

\section{Age Determination}

Two main methods are currently used to measure relative ages: 
{\it (i)} the so-called {\it vertical} method---based on the TO luminosity
with respect to the HB level (Buonanno, Corsi, \& Fusi Pecci
 1989)
 and {\it (ii)} the {\it horizontal} method---based on the accurate
determination of the color difference $\Delta (\bv)_{\rm TO}^{\rm RGB}$ 
between 
the base of the RGB and the TO (VandenBerg, Bolte, \& Stetson 1990). 

Admittedly, our photometry in the [$(\bv)$, $V$] plane is not sufficiently accurate
 to  derive precise  color difference measures down to the TO region.
 In addition,
 the horizontal method is so intrinsically sensitive to small variations 
  in 
colour that even small ($\sim 0.01$~mag) uncertainties in the registration 
of the mean ridge lines could  generate large ($\sim 1$~Gyr),  
``artificial'' differences in the derived ages. 
For this reason, in order to obtain a
relative age for M75, we applied only the vertical method.

In doing this, we perform a comparison with 
respect to another cluster with 
similar metallicity and HB morphology---namely, NGC~1851.
Recent photometry in the [$(V-I)$, $V$] plane
 has been published for NGC1851 by  Bellazzini et al. (2001).
 In Fig.~15 the fiducial lines of NGC\,1851 (from   Bellazzini et al.)
 are overplotted on the   CMD of M75 in the  
  [$(V-I)_{0}$, $V$] plane.

%
\begin{figure*}[t]
  \centerline{\psfig{figure=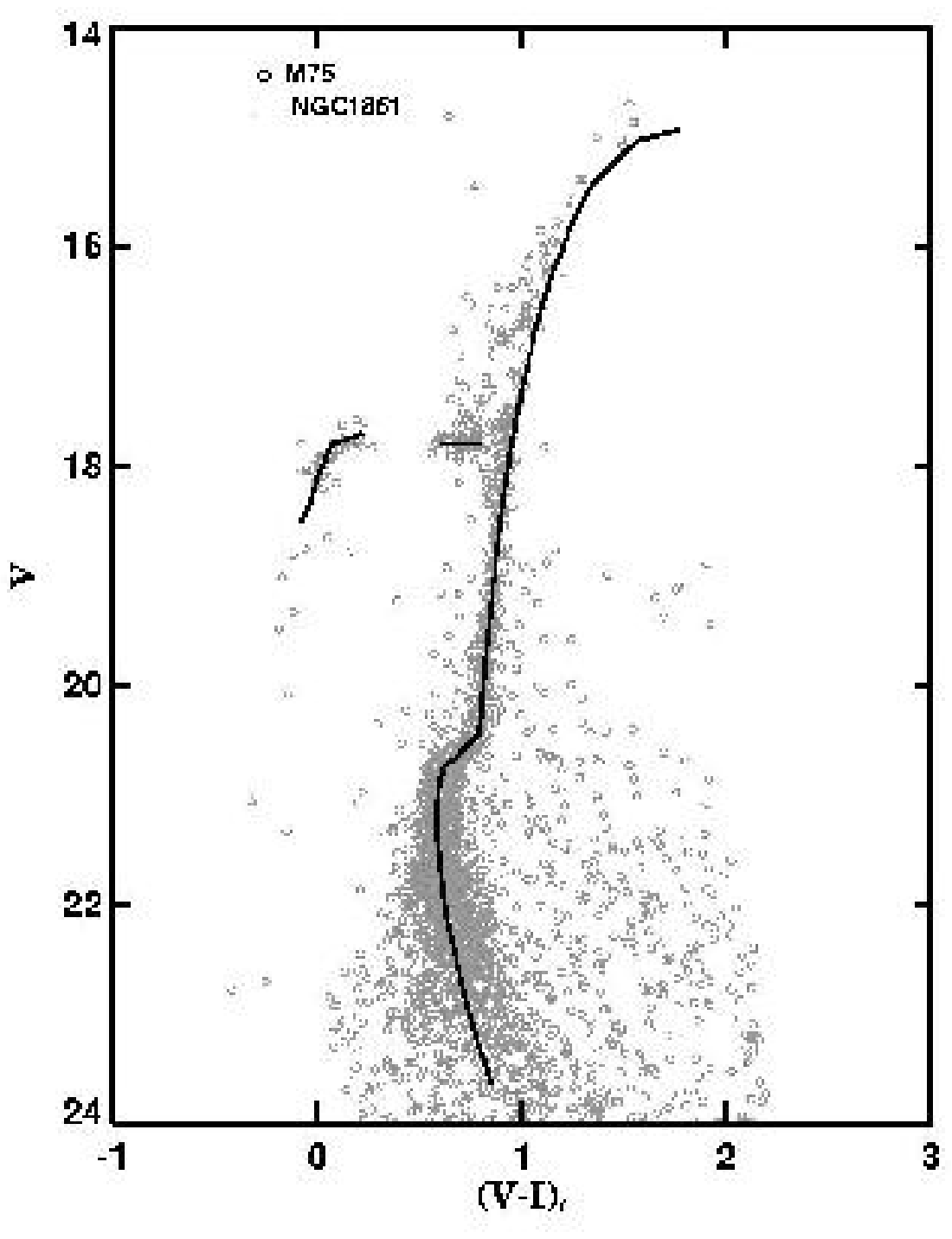}}
  \caption{The [$(V-I)_{0}$, $V$] CMD of  M75 with the 
    fiducial line of NGC\,1851 (from Bellazzini et al. 2001) 
    overplotted  (solid lines).     
    }
      \label{Fig15}
\end{figure*}

As can be seen from Fig.~15, while both the HB and the SGB/TO regions
of the two clusters nicely agree,  
a small difference in colour
      in the brightest  region of the RGB does exist between 
       the two data sets. This small colour residual could in
       principle be due to a small difference in the {\it global} 
       metallicity between the two
       clusters, although the metallicity value derived in 
       \S3.2 for M75 is very similar to that from Ferraro et 
       al. (1999) for NGC~1851. In our opinion, the noted mismatch   
    is likely due to a not completely accounted for ``colour equation''
     in one of the two datasets.
    This effect is generally due to poor coverage in 
  colour of the standard stars.

On the basis of the VandenBerg \& Bell (1985) models, 
Buonanno et al. (1993) obtained the following  ``vertical'' 
relation between the differential age parameter
$\Delta = \Delta_1 V_{\rm TO}^{\rm HB}-
                      \Delta_2 V_{\rm TO}^{\rm HB}$
and the age $t_9$ in Gyr:

\begin{equation}
\Delta \log t_9 = (0.44 + 0.04\,{\rm [Fe/H]})\,\Delta.
\end{equation} 

\noindent In our case, we shall assume that 
subscript  ``1" stands for M75, and  ``2" for NGC\,1851.
For NGC\,1851 we used the data given in Bellazzini et al. 
(2001) to find $\Delta V_{\rm TO}^{\rm HB} = 3.43\pm0.1$. 
Assuming ${\rm [Fe/H]} \simeq -1.05$ for both M75 (\S3.2) 
and NGC~1851 and using the 
value of $\Delta V_{\rm TO}^{\rm HB}$ derived in \S3.1
for M75, we obtain
$\Delta = 3.42 - 3.43 = -0.01\pm0.1$~mag, implying
$\Delta \log t_9 \sim -0.004$ and thus

$$\Delta\,t_9 ({\rm M75} - {\rm NGC\,1851}) \sim 
                               -0.1 \pm 2.0\,\,{\rm Gyr}.$$ 

\noindent This result suggests that M75 is essentially coeval 
with NGC~1851. For recent comparisons of the age of 
NGC~1851 with that of other GCs with similar metallicity, 
including NGC~288 and NGC~362, the reader is referred to 
Rosenberg et al. (1999), VandenBerg (2000), and Bellazzini 
et al. (2001); these papers all seem to indicate that 
NGC~1851 is (slightly) younger than other GCs of similar 
metallicity.

We note that even though the Buonanno et al. (1993) formula, 
Eq.~(1)  above, is not based on the latest evolutionary 
models available in the literature, they are still expected to 
give sufficiently accurate {\em relative} ages, 
under the hypothesis that both clusters have the same chemical 
composition.

\section{Summary and Concluding Remarks}
In this paper, we have provided the first CCD photometry for the distant 
GC M75. We confirm a previous suggestion (Catelan et al. 1998a) that this 
cluster contains two main modes on the HB, and show that the cluster 
actually has a {\em third} mode on the blue tail, clearly separated 
from the bulk of the blue HB by a wide gap, encompassing about 3000~K  
in temperature. In CMDs utilizing broadand $U$ magnitudes, these hotter 
HB stars are found to be clearly inconsistent with canonical evolutionary 
models, showing that the ``Grundahl jump" phenomenon (Grundahl et al. 
1999) can be studied using broadband $U$ besides Str\"omgren $u$. We 
also find a population of centrally concentrated BSS in M75. The 
metallicity inferred for this cluster from our photometry, 
${\rm [Fe/H]} = -1.03\pm 0.17$ in the CG97 scale  
and ${\rm [Fe/H]} = -1.24\pm 0.21$ in the ZW84 scale, 
 is similar to that of the well-known bimodal-HB globular NGC~1851, thus  
enabling a straightforward comparison of their turnoff ages. We find that 
the two clusters have essentially the same age.
This is an interesting result, in view of the fact 
that NGC~1851 itself has been argued to be slightly younger than other 
GCs of similar metallicity (Bellazzini et al. 2001 and references therein). 
However, better photometry is needed to conclusive demonstrate that M75  
too is younger than the average for its metallicity. 

What is the physical reason for the HB multimodality in M75? 
At present, 
this remains unclear. Similar anomalies in HB morphology have been seen 
in other clusters, and the suggested explanations---none of which is 
seemingly capable of accounting for all the peculiarities---usually  
involve environmental effects, stellar rotation, abundance anomalies,  
mass loss in RGB stars, binarity, and even planets (e.g., Sosin et al. 
1997 and references therein). In the case of M75, we are in a bad 
position to analyze most of these factors, given the lack of  
spectroscopic data for the cluster stars. 

However, at least as far 
as the HB bimodality (i.e., relative absence of RR Lyrae stars, compared 
with both blue and red HB stars) is concerned, we can make use of RR  
Lyrae pulsation properties in order to investigate the presence of any  
non-canonical effects that might impact their pulsational properties  
in a significant way. This could at least differentiate processes which 
affect HB morphology by means solely of RGB mass loss from those 
that may affect not only the HB star masses, but also their luminosities 
(Catelan, Sweigart, \& Borissova 1998b). Similarly, useful constraints  
can thereby be placed on the origin of M75's peculiar $R$-ratio. 

With this in mind, we undertook 
an extensive variability survey of M75. In Paper~II we will 
provide a detailed discussion of the variable stars population in this 
cluster, in the hope that this will help us better understand the possible 
origin(s) of its peculiar HB morphology. Accordingly, a more extensive  
discussion of possible explanations for M75's peculiar HB will also be 
deferred to Paper~II.

\begin{acknowledgements}
  Support for M.C. was provided in part
  by NASA through Hubble Fellowship grant HF--01105.01--98A awarded by the 
  Space Telescope Science Institute, which is operated by the Association
  of Universities for Research in Astronomy, Inc., for NASA under
  contract NAS~5--26555. M.C. is grateful to the staff of the Sofia 
  Observatory, where part of this work was carried out, for its hospitality 
  and generous support. F.R.F. acknowledges the financial support of the 
  {\it Agenzia Spaziale Italiana} (ASI) and  the
  {\it Ministero della Ricerca Scientifica e Tecnologica} (MURST).
  This work has been supported in part
  by the U.S. National Science Foundation under grant AST~9986943.
\end{acknowledgements}

\onecolumn

\begin{deluxetable}{cccccc} 
\tablewidth{0pc}
\footnotesize
\tablecaption{Mean Fiducial Lines for M75}
\tablehead{
  \colhead{$V$} & 
  \colhead{$V-I$} &  
\colhead{$V$} & 
  \colhead{$V-I$} &  
    \colhead{$V$} & 
  \colhead{$V-I$}
  }
\startdata
\multicolumn{2}{c}{MS+SGB+RGB}         & \\
14.58  &  1.954    &   20.83  &  0.842    &   17.81      &     0.33     \\
14.86  &  1.750    &   20.90  &  0.831    &   17.88      &     0.26     \\
15.40  &  1.537    &   20.95  &  0.828    &   17.98      &     0.24    \\
15.75  &  1.447    &   21.01  &  0.822    &   18.12      &     0.22     \\
16.25  &  1.338    &   21.12  &  0.819    &   18.80      &     0.12     \\
16.75  &  1.262    &   21.22  &  0.818    &   19.15      &     0.08     \\
17.25  &  1.200    &   21.30  &  0.819    &   19.45      &     0.06     \\
17.75  &  1.149    &   21.40  &  0.819    &   \multicolumn{2}{l}{AGB}   \\
18.25  &  1.110    &   21.50  &  0.822    &   15.84      &     1.32     \\
18.75  &  1.079    &   21.64  &  0.829    &   16.13      &     1.24    \\
19.25  &  1.048    &   21.78 &   0.839    &   16.41      &     1.19    \\
19.72  &  1.026    &   21.93 &   0.850    &   16.72      &     1.12     \\
20.00  &  1.014    &   22.06 &   0.861    &   17.15      &     1.07     \\
20.10  &  1.010    &   22.33 &   0.889    &              &              \\
20.20  &  1.006    &   22.72 &   0.932    &              &              \\
20.30  &  1.003    &   23.05 &   0.978    &              &             \\
20.39  &  0.997    &    \multicolumn{2}{l}{Red HB}       &              \\
20.47  &  0.988    &   17.80      &     1.00&            &             \\
20.54  &  0.970    &   17.80      &     0.90&            &              \\
20.60  &  0.946    &   17.80      &     0.80&            &               \\
20.65  &  0.915    &   17.80      &     0.70&            &                \\
20.70  &  0.884    &     \multicolumn{2}{l}{Blue HB}     &                \\
20.75  &  0.862    &     17.80      &     0.40           &                \\
\enddata
\label{Tab04}
\end{deluxetable}

\begin{deluxetable}{lcl} 
\tablewidth{0pc}
\footnotesize
\tablecaption{M75 Metallicity}
\tablehead{
  \colhead{Parameter} & 
  \colhead{[Fe/H]} 
  }
\startdata
CG97 scale &\\
\tableline
$(\bv)_{0,{\rm g}}$ & $-1.21 \pm 0.22$ \\
$\Delta V_{1.1}$    & $-0.95 \pm 0.16$ \\
$\Delta V_{1.2}$    & $-0.92 \pm 0.14$ \\
$\Delta V_{1.4}$    & $-0.97 \pm 0.12$ \\
$S_{2.5}$    & $-0.74 \pm 0.19$ \\
$S_{2.0}$    & $-0.74 \pm 0.19$ \\
$\Delta V_{\rm HB}^{\rm bump}$    & $-1.10 \pm 0.10$ \\
\tableline
ZW84 scale &\\
\tableline
$(V-B)_{0,{\rm g}}$ & $-1.29 \pm 0.15$\\
$\Delta V_{1.1}$    & $-1.21\pm 0.14$\\
$\Delta V_{1.2}$    & $-1.16 \pm 0.21$ \\
$\Delta V_{\rm HB}^{\rm bump}$    & $-1.30 \pm 0.13$ \\
\enddata
\label{Tab03}
\end{deluxetable}

\begin{deluxetable}{lccc} 
\tablewidth{0pc}
\footnotesize
\tablecaption{Ratio between HB and RGB Stars in M75}
\tablehead{
  \colhead{Parameter} & 
  \colhead{Observed} & 
  \colhead{Model, Unimodal} & 
  \colhead{Model, Bimodal}  
  }
\startdata
 $R_{\langle M_V\rangle}$         &   $1.75 \pm 0.20$ & $1.52 \pm 0.20$ & $1.52 \pm 0.20$  \\
 $R_{\langle M_V\rangle+1.0}$     &   $0.89 \pm 0.12$ & $0.73 \pm 0.08$ & $0.73 \pm 0.08$  \\
 $R_{\langle M_V\rangle+1.5}$     &   $0.80 \pm 0.09$ & $0.51 \pm 0.05$ & $0.51 \pm 0.05$  \\
\enddata
\label{Tab0z}
\end{deluxetable}

\begin{deluxetable}{lccc} 
\tablewidth{0pc}
\footnotesize
\tablecaption{HB Morphology Parameters for M75}
\tablehead{
  \colhead{Parameter} & 
  \colhead{Observed} & 
  \colhead{Model, Unimodal} & 
  \colhead{Model, Bimodal}  
  }
\startdata
$B:\,V:\,R$      &  $0.38:\,0.16:\,0.45$ & $0.369:\,0.164:\,0.467$ & $0.374:\,0.162:\,0.463$ \\ 
              &  &$(0.040) \, (0.030)\, (0.040)$ &$(0.040)\, (0.030)\, (0.039)$ \\ 
$B/(B+R)$        &  $+0.46\pm 0.05$      &  $+0.442\pm0.045$  & $+0.447\pm 0.044$ \\
$(B-R)/(B+V+R)$  &  $-0.07\pm 0.03$      &  $-0.097\pm0.075$  & $-0.089\pm 0.073$ \\ 
$(B2-R)/(B+V+R)$ &  $-0.29\pm 0.05$      &  $-0.251\pm0.064$  & $-0.332\pm 0.056$ \\ 
$B2'/(B+V+R)$    &  $+0.94\pm 0.11$      &  $+0.820\pm0.032$  & $+0.924\pm 0.022$ \\ 
$B7'/(B+V+R)$    &  $+0.15\pm 0.05$      &  $+0.099\pm0.025$  & $+0.150\pm 0.029$ \\ 
$B0'/B$          &  $+0.65\pm 0.13$      &  $+0.449\pm0.068$  & $+0.705\pm 0.063$ \\
\enddata
\label{TabHB}
\end{deluxetable}

\begin{deluxetable}{ccccccc} 
\tablewidth{0pc}
\footnotesize
\tablecaption{BSS Candidates}
\tablehead{
  \colhead{$x$} & 
  \colhead{$y$} &
  \colhead{$U$} & 
  \colhead{$B$} & 
  \colhead{$V$} & 
  \colhead{$I$} & 
  \colhead{$r(\arcmin)$} 
  }
\startdata
   858.487& 928.874&  20.726&   21.033&   20.234&   19.765&   1.197\\
 1569.31& 509.059&  20.706&   20.695&   20.234&   19.595&   3.174\\
 856.750& 1025.86&  19.966&   20.005&   19.544&   19.145&   1.083\\
642.167& 1130.38&  20.616&   20.649&   20.224&   19.645&   2.060\\
1043.11& 561.039&  20.776&   21.033&   20.394&   19.565&   2.182\\
1221.19& 994.192&  20.616&   20.954&   20.244&   19.415&   0.591\\
1029.49& 361.2  &  20.586&   20.739&   20.234&   19.455&   3.069\\
393.541& 354.62 &  20.696&   20.779&   20.354&   19.855&   4.394\\
969.387& 756.66 &  20.446&   0     &   20.114&   19.515&   1.423\\
1380.32& 889.383&  20.786&   20.961&   20.464&   19.705&   1.432\\
664.147& 1643.63&  20.406&   20.527&   20.084&   19.295&   3.264\\
1245.00& 1180.94&  20.546&   0     &   20.234&   19.515&   0.866\\
903.636& 1184.78&  20.696&   20.855&   20.394&   19.565&   1.055\\
908.318& 665.808&  20.696&   0     &   20.404&   19.575&   1.903\\
1211.07& 1287.94&  20.726&   20.744&   20.434&   19.595&   1.164\\
1416.58& 662.593&  20.606&   0     &   20.334&   19.405&   2.218\\
1342.55& 1132.68&  20.556&   20.666&   20.294&   19.415&   1.136\\
1078.26& 1238.95&  20.526&   20.629&   20.284&   19.465&   0.843\\
1117.41& 1278.94&  20.286&   20.429&   20.084&   19.285&   1.017\\
1200.79& 512.574&  20.546&   20.699&   20.354&   19.495&   2.424\\
668.343& 1009.46&  20.236&   20.31 &   20.054&   19.155&   1.922\\
1272.36& 1202.11&  20.626&   0     &   20.474&   19.665&   1.019\\
1203.81& 1231.74&  20.616&   20.748&   20.474&   19.765&   0.927\\
1092.07& 1312.79&  20.346&   20.39 &   20.214&   19.365&   1.165\\
972.800& 771.894&  20.526&   0     &   20.414&   19.595&   1.355\\
1239.57& 852.423&  20.066&   0     &   20.234&   19.485&   1.072\\
\enddata
\label{Tab02}
\end{deluxetable}

\end{document}